\documentclass[conference]{IEEEtran}
\usepackage{fancyhdr}
\IEEEoverridecommandlockouts

\usepackage[T1]{fontenc} \usepackage{graphicx} \usepackage{booktabs} \usepackage{amsmath,amssymb} \usepackage{xcolor} \usepackage{cite} \usepackage{url} \usepackage{tikz} \usetikzlibrary{arrows.meta,positioning,shapes.geometric,fit,backgrounds,calc,matrix,patterns} \usepackage[hidelinks]{hyperref}   % load last

\newif\ifblind \blindfalse

\ifblind
  \newcommand{\sys}{\textsc{SupportLoop}}
  \newcommand{\Fac}{the facility}          % mid-sentence
  \newcommand{\FacLong}{a U.S.\ Department of Energy leadership computing facility}
  
  \newcommand{\FacAdj}{facility}
  \newcommand{\SysA}{System~A}             % NVIDIA / CUDA flagship
  \newcommand{\SysB}{System~B}             % Intel oneAPI / SYCL flagship
  \newcommand{\Gw}{the in-facility inference gateway}

\else
  \newcommand{\sys}{AskALCF}
  \newcommand{\Fac}{the ALCF}
  
  \newcommand{\FacLong}{the Argonne Leadership Computing Facility (ALCF)}
  
  \newcommand{\FacAdj}{ALCF}
  \newcommand{\SysA}{Polaris}
  \newcommand{\SysB}{Aurora}
  \newcommand{\Gw}{the ALCF Argo inference gateway}

\fi

\newcommand{\code}[1]{\texttt{#1}}
\newcommand{\stRun}{\textcolor{green!45!black}{$\blacksquare$}}
\newcommand{\stPart}{\textcolor{orange!80!black}{$\square$}}
\newcommand{\stFut}{\textcolor{black!45}{$\square$}}
\colorlet{rtNoFailS}{green!45!black}  \colorlet{rtNoFailF}{rtNoFailS!28}
\colorlet{rtRetrS}  {blue!55!black}   \colorlet{rtRetrF}  {rtRetrS!26}
\colorlet{rtKnowS}  {orange!65!black} \colorlet{rtKnowF}  {rtKnowS!32}
\colorlet{rtPromptS}{violet!60!black} \colorlet{rtPromptF}{rtPromptS!30}
\colorlet{rtModelS} {violet!60!black} \colorlet{rtModelF} {rtModelS!11}
\colorlet{rtToolsS} {black!60}        \colorlet{rtToolsF} {rtToolsS!45}
\colorlet{rtUnkS}   {black!45}        \colorlet{rtUnkF}   {rtUnkS!20}
\newcommand{\origintag}[1]{{\scriptsize\bfseries\color{blue!60!black}#1}}
\newcommand{\contribtag}[1]{{\scriptsize\bfseries\color{orange!60!black}#1}}
\newcommand{\bul}{\strut$\cdot$\ }
\newcommand{\HZ}[1]{}
\newif\ifpendingvisible \pendingvisibletrue

\begin{document}

% ---------------------------------------------------------------------
% TITLE
% ---------------------------------------------------------------------
\title{\sys: A Self-Improving Agentic System for\\
        Intelligent HPC User Support}

\ifblind
\author{\IEEEauthorblockN{Anonymous Author(s)}
\IEEEauthorblockA{\textit{Affiliation withheld for double-blind review}}}
\else
% Camera-ready author list and affiliations confirmed against the SC26
% submission metadata. Argonne divisions are intentionally omitted.
\newcommand{\scauthor}[4]{%
\begin{minipage}[t]{0.31\textwidth}
\centering
{\normalsize #1\par}
\vspace{0.25ex}
{\footnotesize\itshape
#2\par
#3\par
\href{mailto:#4}{#4}\par}
\vspace{4ex}
\end{minipage}}
\newcommand{\scauthordual}[6]{%
\begin{minipage}[t]{0.31\textwidth}
\centering
{\normalsize #1\par}
\vspace{0.25ex}
{\footnotesize\itshape
#2\par
#3\par
#4\par
#5\par
\href{mailto:#6}{#6}\par}
\vspace{4ex}
\end{minipage}}
\author{%
\begin{tabular}{@{}c@{\hspace{0.015\textwidth}}c@{\hspace{0.015\textwidth}}c@{}}
\scauthor{Jingyan Jiang\textsuperscript{*}}
  {Argonne National Laboratory}{Lemont, IL, USA}{jingyan.jiang@anl.gov} &
\scauthordual{Akhilesh Bondapalli\textsuperscript{*}}
  {Clemson University}{Clemson, SC, USA}
  {Argonne National Laboratory}{Lemont, IL, USA}{abondap@clemson.edu} &
\scauthordual{Krishna Rauniyar}
  {University of Louisiana at Lafayette}{Lafayette, LA, USA}
  {Argonne National Laboratory}{Lemont, IL, USA}{krishna.rauniyar1@louisiana.edu}
\\
\scauthor{Murat Keceli}
  {Argonne National Laboratory}{Lemont, IL, USA}{keceli@anl.gov} &
\scauthor{Yasaman Ghadar}
  {Argonne National Laboratory}{Lemont, IL, USA}{ghadar@anl.gov} &
\scauthor{Haritha Siddabathuni Som}
  {Argonne National Laboratory}{Lemont, IL, USA}{haritha@anl.gov}
\\
\scauthor{Venkatram Vishwanath}
  {Argonne National Laboratory}{Lemont, IL, USA}{venkat@anl.gov} &
\scauthor{Michael E. Papka}
  {Argonne National Laboratory}{Lemont, IL, USA}{papka@anl.gov} &
\scauthor{Rong Ge}
  {Clemson University}{Clemson, SC, USA}{rge@clemson.edu}
\\
& \scauthor{Huihuo Zheng\textsuperscript{\dag}}
  {Argonne National Laboratory}{Lemont, IL, USA}{huihuo.zheng@anl.gov} &
\end{tabular}\vspace{-4ex}}
\fi

\maketitle
\ifblind\else
\begingroup
\renewcommand{\thefootnote}{\fnsymbol{footnote}}
\footnotetext[1]{Joint first authors.}
\footnotetext[2]{Corresponding author.}
\endgroup
\fi
\thispagestyle{fancy}
\lhead{}
\rhead{}
\chead{}
\lfoot{\footnotesize{
SC26 Workshops, November 15-20, 2026, Chicago, Illinois, USA
\newline 979-8-3195-1221-5/26/\$31.00 \copyright 2026 IEEE}}
\rfoot{}
\cfoot{}
\renewcommand{\headrulewidth}{0pt}
\renewcommand{\footrulewidth}{0pt}

\begin{abstract}
AI-based question-answering systems are increasingly used for user support, including at
high-performance computing (HPC) facilities. However, existing systems lack a systematic,
deployment-oriented approach for evaluating final-answer quality and attributing failures
to specific components. They also remain largely static after deployment, apart from
periodic knowledge-base updates.

We present \sys, an agentic log-analysis and improvement pipeline for HPC user support.
\sys{} evaluates production interactions at a fine-grained level, identifies whether
failures arise from missing knowledge, retrieval, prompting, model behavior, or tool
execution, and routes each failure to an actionable repair. Its diagnostic core combines
answer-quality assessment with a two-scope sufficiency test that distinguishes information
missing from the documentation corpus from information present but not retrieved. We
further introduce a controlled harness-training process that uses expert references, null
edits, and explicit ship, revert, or decline decisions to improve the diagnostic pipeline
reliably.

Across 2{,}174 support turns, \sys{} reduced unattributable failures from 41.4\% to
14.2\%, while harness training improved expert agreement from 108/150 to 123/150 turns.
These results demonstrate a governed form of self-evolution in which production evidence
drives targeted system improvements. By turning production interactions into fine-grained
diagnoses and actionable repairs, \sys{} provides a systematic foundation for
continuously improving deployed AI support systems beyond knowledge-base updates alone.
\end{abstract}


% =====================================================================

% =====================================================================
% BODY --- sections split into ./sections/
% =====================================================================
\section{Introduction}
\label{sec:intro}
% =====================================================================
Leadership computing facilities support thousands of users across systems whose
hardware, software and operating procedures change continuously. New partitions
come online, programming environments evolve, and compiler, driver and library
versions change between maintenance cycles. At the same time, the information
users need is distributed across facility guides, vendor manuals, software
documentation, release notes and separate web portals. Finding the right
information is therefore difficult even when it has been documented. Together,
rapid change, fragmented documentation and growing demand create a persistent
support burden for both users and facility staff.

Retrieval-augmented generation (RAG) has become a promising way to make this
support more accessible. Systems such as AskHPC~\cite{askhpc2025} and
chatHPC~\cite{chathpc2025}, HyCE's
HPC-oriented RAG architecture~\cite{miyashita2025hyce}, and recent
cross-facility studies~\cite{zheng2026comparative} show that language models can
answer technical questions using facility documentation. These systems
represent an important advance over static documentation search: users can ask
questions in natural language and receive an immediate, context-aware response.

However, current systems remain limited in two important ways. First, detecting
a poor answer does not reveal how to fix it. The same visible error may result
from missing documentation, failed retrieval, an inadequate prompt, model
behavior, or a broken tool call. These causes belong to different teams and
require different repairs. A single quality score can identify that an answer
is weak, but it cannot identify who should act or what they should change.

Second, deployed assistants make little systematic use of the evidence they
accumulate. Every interaction records a real user need, the material retrieved
for it, and the answer produced. Support tickets provide another source of
operational knowledge: they capture problems that may require an existing
answer, reproduction on \FacAdj{} systems, or intervention by a system owner.
Yet chatbot logs and support tickets are typically handled as separate records
rather than as evidence for continuously improving the support system.

We present \sys{}, an agentic support and improvement framework that connects
these two channels. Interactive questions are answered by a deployed RAG
service, while a diagnostic pipeline analyzes the resulting logs and routes
failures to the component that can address them. Support tickets are classified
by the type of resolution they require, and responses are drafted for expert
review. A shared knowledge-management process collects lessons from both
channels and turns them into reviewed updates to the documentation corpus.
Together, these four components allow the system to learn from production use,
improve its diagnostic rules and prompts, and accumulate reusable knowledge. We
refer to this controlled process as \emph{governed self-evolution}: the system
proposes and evaluates improvements, while human experts retain authority over
responses and deployed changes.

The central idea is to distinguish an answer's visible symptom from its
underlying cause. In particular, \sys{} asks two separate questions: whether the
retrieved passages were sufficient to answer the user, and whether the required
information existed anywhere in the documentation corpus. This distinction,
inspired by sufficient-context evaluation~\cite{sufficientcontext2024},
separates missing knowledge from failed retrieval. Combined with evidence about
the answer, model and tool execution, it allows the system to route a failure
to one of five component owners rather than returning only a quality score. An
agent judge independently audits the route, while \FacAdj{} experts provide the
final reference for disputed or uncertain cases.

This changes the role of an AI support assistant. Instead of treating each
interaction as an isolated question-answer exchange, \sys{} treats production
use as evidence for improving the support process itself. The result is a
governed loop from user request, to diagnosis, to repair, to reviewed knowledge.
The significance is not simply that the assistant can produce another answer;
it is that the system can identify an actionable repair, direct it to the right
owner, and preserve the resulting lesson for future users.

We evaluate this approach on 2{,}174 production support turns. The diagnostic
pipeline reduces failures that cannot be assigned to an owner from 41.4\% to
14.2\%. A human-guided improvement process increases agreement with a frozen
three-expert reference from 108/150 to 123/150 turns. Controlled repair
experiments further show that the resulting diagnoses can guide effective
prompt and documentation changes. Together, these results demonstrate that
production support evidence can be converted into measurable and actionable
system improvements.

\noindent\textbf{Contributions.} This paper makes four contributions:

\begin{itemize}
  \item \textbf{An actionable diagnostic for AI-based HPC support}
  (\S\ref{sec:architecture}, \S\ref{sec:diagnostic}). The diagnostic
  distinguishes failures caused by knowledge, retrieval, prompting, model
  behavior and tool execution, and routes each failure to the component owner
  who can address it.

  \item \textbf{A governed agent judge for production diagnosis}
  (\S\ref{sec:judge}, \S\ref{sec:overturn}). The judge audits a deterministic
  rule tree rather than replacing it, names the evidence behind a disagreement,
  and can change a route only under explicitly evaluated rules.

  \item \textbf{A human-guided method for improving the diagnostic pipeline}
  (\S\ref{sec:harness}). We treat the rules, prompts and judge policy---rather
  than model weights---as the object of improvement, and evaluate proposed
  changes against fixed expert references and explicit controls.

  \item \textbf{An integrated system for governed self-evolution}
  (\S\ref{sec:architecture}, \S\ref{sec:production}). By combining question
  answering, log analysis, ticket resolution and knowledge management, \sys{}
  converts production experience into proposed repairs, evaluates their effects,
  and preserves approved lessons for future requests. Controlled prompt and
  documentation experiments provide evidence that this loop can produce
  effective improvements while experts retain authority over consequential
  changes.
\end{itemize}

\noindent\textbf{Scope.} The improvement loop is implemented and evaluated
through expert review. We diagnose production interactions, author targeted
repairs, and measure their effects in controlled experiments. No
CODEOWNERS-approved knowledge change has yet been merged into the deployed
corpus and re-indexed. We therefore report measured offline improvements, not a
claim that the production system has already modified itself autonomously.

\section{Background and Related Work}
% =====================================================================

\subsection{HPC Assistants and RAG Diagnosis}
\label{sec:related-hpc}
\label{sec:related-rag}

AskHPC and chatHPC~\cite{askhpc2025,chathpc2025} answer user questions from
facility documentation, with chatHPC also supporting script generation.
HyCE~\cite{miyashita2025hyce} extends retrieval to HPC-specific operational
data, while cross-facility studies show that support behavior does not transfer
cleanly between sites~\cite{zheng2026comparative}. These systems demonstrate
the value of language models for HPC support, but their principal output is an
answer rather than a diagnosis or repair path.

Existing RAG evaluations measure complementary aspects of retrieval and
generation. RAGAS~\cite{ragas2023} and RAGChecker~\cite{ragchecker2024} provide
multi-dimensional evaluation, while citation-based methods assess whether
claims are supported by retrieved evidence~\cite{alce2023}. The Sufficient
Context test asks whether the passages returned for a question contain enough
information to answer it~\cite{sufficientcontext2024}. \sys{} extends this idea
to two scopes: the passages returned in production and the complete corpus.
The difference distinguishes a retrieval failure from a documentation gap and
therefore identifies the component owner who can act.

Other work uses relevance or sufficiency judgments to adapt retrieval during a
turn~\cite{crag2024,selfrag2023}, and taxonomies describe recurring RAG
engineering failures~\cite{barnett2024sevenfp}. These methods improve an answer
or characterize a failure, but do not connect production evidence to a facility
support workflow. \textbf{Our contribution is an operational diagnosis:} the
two-scope sufficiency test is combined with answer and tool evidence to assign
each production turn to a component-level route and a corresponding repair
owner.

\subsection{Agentic Evaluation and Governance}
\label{sec:related-judges}

Agent-as-a-Judge systems use tools to acquire evidence before producing a
verdict~\cite{zhuge2024agentasjudge,you2026agentasjudgesurvey}. This capability
is useful for failure attribution because an agent can inspect evidence that
was absent from the original interaction. However, LLM judges remain sensitive
to position, self-preference, and other systematic
biases~\cite{zheng2023judging,panickssery2024selfrecog}; self-reported confidence
is also an unreliable indicator of correctness~\cite{overconfidence2025}.
Executable software agents introduce a related governance problem: their
localize--repair--validate workflows require bounded tools and explicit safety
controls~\cite{codeact2024,toolemu2024}.

Agentic failure-attribution systems similarly use tools to localize faults in
execution traces~\cite{zhu2026safari,li2026errorprobe}. Existing governance
strategies commonly rely on confidence thresholds, escalation, or multiple
judges~\cite{trustorescalate2024,verga2024poll}. These strategies are difficult
to apply when confidence is not calibrated to correctness.

\sys{} therefore combines deterministic and agentic evaluation. An ordered rule
tree assigns the initial failure route, an agent judge independently probes the
corpus, and a rule-conditioned gate limits when the judge may change the route.
Human review remains mandatory for ticket responses, knowledge changes, and
generated repairs. \textbf{Our contribution is the division of authority:} the
judge may overturn only decisions based on evidence it can independently test,
while experts retain control over every production-facing action.

\subsection{Human-Guided Self-Improvement}
\label{sec:related-selfimprove}

Self-improving agents commonly update an internal memory or skill library from
their own traces~\cite{wang2024voyager,shinn2023reflexion}, while systems such
as DSPy and GEPA optimize prompts or language-model programs against an
evaluation objective~\cite{dspy2023,agrawal2025gepa}. Recent systems also learn
skills from operational support traces or maintain documentation through
agentic feedback~\cite{liu2026skillforge,badrinarayan2026docsync}. These
approaches differ primarily in what they modify and who determines whether a
change is acceptable.

Pipeline optimizers can also propagate natural-language feedback through
prompts and agent trajectories~\cite{textgrad2024,cheng2024trace}. However,
small prompt changes can interact with nondeterministic generation and produce
apparent gains that do not replicate~\cite{atil2024nondeterminism,
zhang2026coinflip}. A self-improvement claim therefore requires both a stable
reference and controls for run-to-run variation.

In \sys{}, domain experts provide both corrected routes and written rationales.
These observations motivate bounded changes to diagnostic rules, judge
instructions, and \FacAdj{} knowledge. Each change is evaluated against a
frozen, session-disjoint expert reference before acceptance. The resulting
process improves the surrounding harness rather than model weights, while
human reviewers retain authority over every production-facing change.
\textbf{Our contribution is a governed self-evolution protocol} that turns
expert explanations into testable harness changes, records accepted and
rejected arms, and promotes a change only when its target cases improve and its
controls remain stable.

\section{System Design and Implementation}
\label{sec:architecture}
% =====================================================================

At \FacLong{}, users reach \sys{} through two channels. Short how-to questions arrive
through web chat or MCP~\cite{mcp2025} and need an immediate answer. The service
covers \SysA{} and \SysB{} and sends model requests through \Gw{}. Problem
reports arrive by email and become support tickets. A ticket may be answered
from existing knowledge, sent to an agent to reproduce and repair the problem
on real HPC resources, or routed to the team responsible for a system-level
fix.

These needs motivate four pillars. Pillar~1 answers chatbot questions, and
Pillar~3 analyzes the resulting logs to identify failures and improve the
service. Pillar~4 triages support tickets and directs each one to an appropriate
resolution path. Pillar~2 collects lessons from both channels and turns them
into reviewed knowledge updates. Together, the four pillars connect immediate
assistance, failure analysis, ticket resolution, and knowledge curation in a
single human-governed improvement loop. Because a ticket's \emph{complete}
flag does not reliably indicate resolution (\S\ref{sec:distribution}),
Pillar~4 re-triages every ticket from its content.

\begin{figure*}[t]
\centering
\begin{tikzpicture}[
  font=\scriptsize,
  inherited pillar/.style={
    draw=blue!55!black, thick, rounded corners=2pt,
    minimum width=3.0cm, minimum height=2.7cm,
    align=left, inner sep=4pt, text width=2.72cm,
    fill=blue!5
  },
  contrib pillar/.style={
    draw=orange!60!black, line width=0.9pt, rounded corners=2pt,
    minimum width=3.0cm, minimum height=2.7cm,
    align=left, inner sep=4pt, text width=2.72cm,
    fill=orange!5
  },
  gate/.style={
    draw=red!55!black, thick, dashed, rounded corners=2pt,
    minimum width=1.6cm, minimum height=1.7cm,
    align=center, inner sep=3pt, text width=1.39cm,
    fill=red!5
  },
  persona/.style={
    draw=black!35, rounded corners=1pt, inner sep=3pt,
    font=\scriptsize\itshape, align=center, fill=black!5,
    minimum width=1.7cm
  },
  sink/.style={
    draw=black!35, rounded corners=1pt, inner sep=3pt,
    font=\scriptsize\itshape, align=center, fill=black!3,
    minimum width=1.8cm
  },
  flow arrow/.style={->, >={Latex[length=1.8mm]}, thick, black!75},
  side arrow/.style={->, >={Latex[length=1.8mm]}, thick, black!55, dashed},
  edge label/.style={font=\scriptsize, align=center},
  edge num/.style={font=\scriptsize\bfseries, black!90}
]
\node[inherited pillar] (p1) at (0,0) {%
  \textbf{Pillar 1 --- Q\&A}\\
  \origintag{RAG SERVICE}\\[2pt]
  \stRun~Milvus (multi-facility)\\
  \stRun~3 LLM backends\\
  \stRun~web chat $+$ MCP
};
\node[contrib pillar, right=0.5cm of p1] (p3) {%
  \textbf{Pillar 3 --- Log Analysis}\\
  \contribtag{NEW}\\[2pt]
  \stRun~Diagnostic core\\
  \phantom{\stRun~}(\S\ref{sec:diagnostic})\\
  \stRun~answer $+$ retrieval\\
  \phantom{\stRun~}scoring $\to$ 7 routes
};
\node[contrib pillar, right=0.5cm of p3] (p4) {%
  \textbf{Pillar 4 --- Support Tickets}\\
  \contribtag{NEW}\\[2pt]
  \stRun~ingest $+$ classify\\
  \stRun~route to a resolution path\\
  \stRun~drafted email replies
};
\node[contrib pillar, right=0.5cm of p4] (p2) {%
  \textbf{Pillar 2 --- Know.\ Mgmt}\\
  \contribtag{NEW}\\[2pt]
  \stRun~\texttt{gap-miner}\\
  \stRun~\texttt{know-writer}\\
  \stRun~\texttt{reviewer}\\
  \stPart~open PR (dry-run ok)\\
  \stFut~re-index on merge
};
\node[gate, right=0.5cm of p2] (gate) {%
  \textbf{HUMAN}\\
  \textbf{MERGE}\\
  \textbf{GATE}\\[1pt]
  {\tiny CODEOWNERS}
};

\node[persona, above=0.4cm of p1] (user)
  {end user + agent client\\{\tiny (web chat / MCP)}};
\node[persona, above=0.4cm of p4] (ticket)
  {ticket originator\\{\tiny (ticket system / email / chat)}};
\node[persona, above=0.9cm of gate] (maintainer) {maintainer};
\node[sink, below=0.5cm of p4] (enduser)
  {$\to$ support staff\\{\tiny (drafted reply)}};
\node[sink, below=0.4cm of p3] (extown)
  {$\to$ the team that owns it\\{\tiny (route $\ne$ knowledge)}};

\begin{scope}[on background layer]
  \draw[fill=blue!5, draw=none, rounded corners=2pt]
    let \p1 = (p1.west), \p2 = (p1.east) in
    (\x1-0.20cm, -3.30) rectangle (\x2+0.20cm, 3.50);
  \draw[fill=orange!3, draw=none, rounded corners=2pt]
    let \p1 = (p3.west), \p2 = (p4.east) in
    (\x1-0.15cm, -3.30) rectangle (\x2+0.15cm, 3.50);
  \draw[fill=orange!8, draw=none, rounded corners=2pt]
    let \p1 = (p2.west), \p2 = (p2.east) in
    (\x1-0.15cm, -3.30) rectangle (\x2+0.15cm, 3.50);
  \draw[fill=red!7, draw=none, rounded corners=2pt]
    let \p1 = (gate.west), \p2 = (gate.east) in
    (\x1-0.15cm, -3.30) rectangle (\x2+0.15cm, 3.50);
\end{scope}

\node[
  draw=blue!55!black, fill=blue!12, rounded corners=2pt, thin,
  minimum height=0.72cm, minimum width=3.0cm, align=center, inner sep=2pt
] at ($(p1)+(0,3.10)$)
  {\textbf{\textsc{run}}\\[-1pt]{\itshape\tiny per query (sec)}};
\node[
  draw=orange!55!black, fill=orange!12, rounded corners=2pt, thin,
  minimum height=0.72cm, minimum width=6.5cm, align=center, inner sep=2pt
] at ($(p3)!0.5!(p4)+(0,3.10)$)
  {\textbf{\textsc{watch}}\\[-1pt]{\itshape\tiny P3 diagnoses traces (min), P4 works tickets (min--hr)}};
\node[
  draw=orange!60!black, fill=orange!22, rounded corners=2pt, thin,
  minimum height=0.72cm, minimum width=3.0cm, align=center, inner sep=2pt
] at ($(p2)+(0,3.10)$)
  {\textbf{\textsc{improve}}\\[-1pt]{\itshape\tiny per PR (days)}};
\node[
  draw=red!55!black, fill=red!14, rounded corners=2pt, thin,
  minimum height=0.72cm, minimum width=1.6cm, align=center, inner sep=2pt
] at ($(gate)+(0,3.10)$)
  {\textbf{\textsc{gate}}\\[-1pt]{\itshape\tiny human review}};

\draw[flow arrow] (user.south) -- (p1.north)
  node[edge num, midway, right, xshift=2pt] {(1)};
\draw[flow arrow] (ticket.south) -- (p4.north)
  node[edge num, midway, right, xshift=2pt] {(0)};
\draw[flow arrow] (p1.east) -- (p3.west)
  node[edge num, midway, above=1pt] {(2)};

\coordinate (a3bot-l) at ($(p3.south east)+(0,-1.55)$);
\coordinate (a3bot-r) at ($(p2.south west)+(0,-1.55)$);
\draw[flow arrow] (p3.south east) -- (a3bot-l);
\draw[flow arrow] (a3bot-l) -- (a3bot-r)
  node[edge num, midway, below=1pt] {(3)};
\draw[flow arrow] (a3bot-r) -- (p2.south west);

\draw[side arrow] (p3.south) -- (extown.north)
  node[edge num, midway, right, xshift=2pt] {(3b)}
  node[edge label, midway, left, xshift=-2pt, font=\tiny] {other 4};
\draw[->, >={Latex[length=1.8mm]}, line width=1.2pt, orange!60!black]
  (p4.east) -- (p2.west)
  node[edge num, midway, above=1pt] {(5)};
\draw[flow arrow] (p4.south) -- (enduser.north)
  node[edge num, midway, right, xshift=2pt] {(4)};
\draw[flow arrow] (p2.east) -- (gate.west)
  node[edge num, midway, above=1pt] {(6)};
\draw[side arrow] (maintainer.south) -- (gate.north)
  node[edge label, midway, right, xshift=2pt] {reviews};
\draw[flow arrow] (gate.south) |- ($(enduser.south)+(0,-0.7)$) -| (p1.south);
\node[edge num] at ($(gate.south)+(0.35,-0.35)$) {(7)};
\node[edge label] at ($(enduser.south)+(-2.5,-0.85)$)
  {merged $\to$ vector re-index \stFut};

\node[edge label, anchor=north, font=\scriptsize] at ($(p4.south)+(0,-2.95)$)
  {\stRun~runs today \quad
   \stPart~built, not exercised in production \quad
   \stFut~designed and coded, never run};
\end{tikzpicture}
\caption{\sys{} connects two support channels to a governed improvement loop.
A user query enters Pillar~1 (1), whose trace is analyzed by Pillar~3 (2).
Confirmed knowledge gaps go to Pillar~2 (3), while other failures go to their
component owners (3b). A ticket enters Pillar~4 (0), which drafts a reply for
staff review (4) and sends reusable fixes to Pillar~2 (5). Pillar~2 opens a pull
request (6); after human approval, the merge trigger re-indexes the corpus (7).
That final trigger has not yet fired.}
\label{fig:pillars}
\end{figure*}
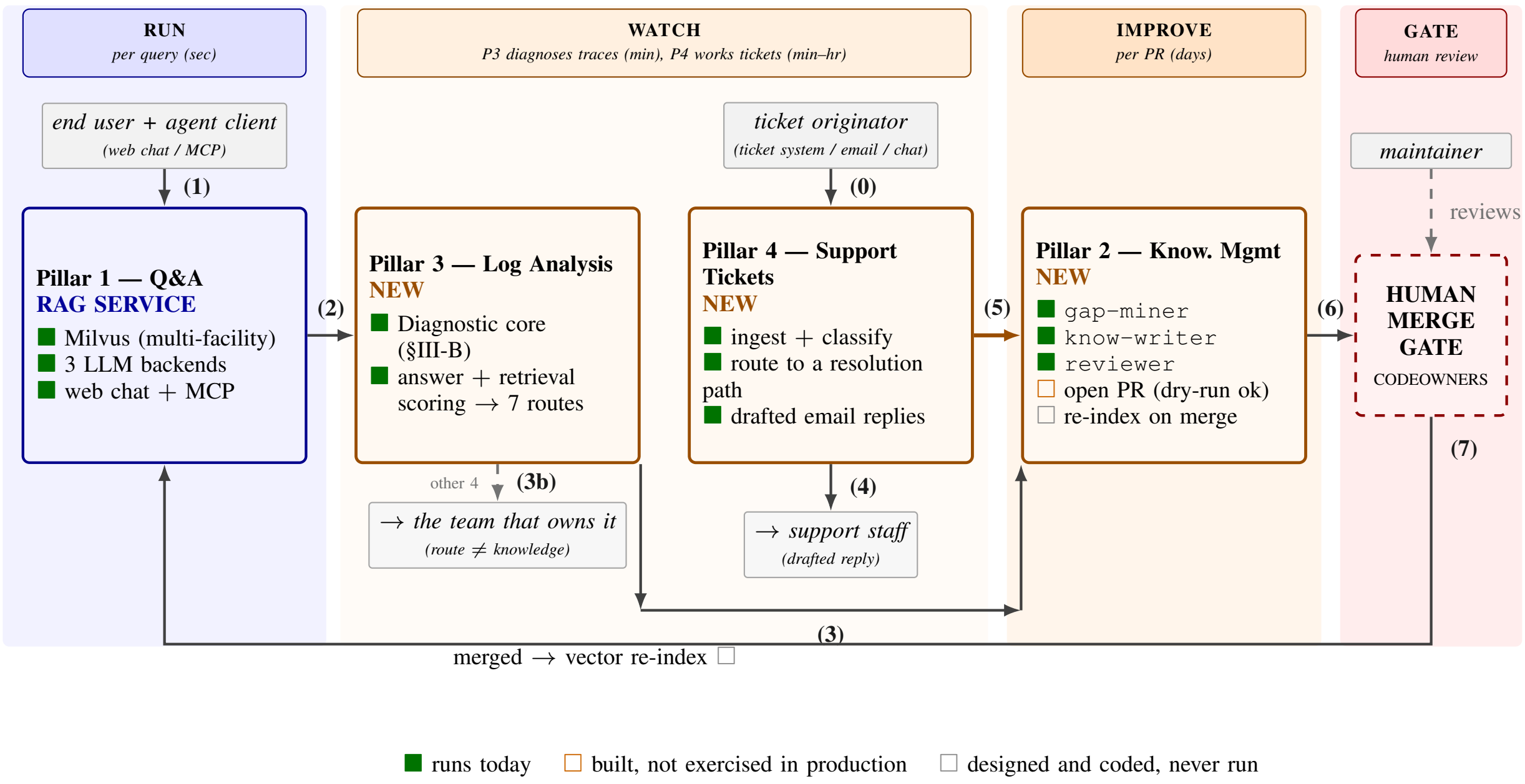

\subsection{The Four Subsystems}
\label{sec:pillars}

\textbf{Pillar~1 --- Q\&A service.} This deployed RAG service indexes a curated
corpus of facility user guides, vendor documentation and other operational
materials in Milvus~\cite{milvus2021}. For each question, it retrieves relevant
passages and sends them to a language model as context for answer generation.
It logs the question, answer, retrieved passages, tool calls and retrieval
score. Pillar~3 uses this record to evaluate what the production system
actually saw.

\textbf{Pillar~3 --- Log analysis.} Pillar~3 examines each failing Pillar~1
turn and assigns one of five causes: \code{knowledge}, \code{retrieval},
\code{prompt}, \code{model} or \code{tools\_mcp}. A \code{knowledge} failure
means that the required information is absent from the indexed corpus, whereas
a \code{retrieval} failure means that the information exists but was not
returned to the model. A \code{prompt} failure occurs when the system
instructions induce inappropriate behavior, such as unnecessary clarification.
A \code{model} failure occurs when sufficient context and appropriate
instructions are available but the model produces an incorrect or unsupported
answer. A \code{tools\_mcp} failure arises when an external tool, MCP service or
its invocation fails. We use \code{no\_failure} when the evidence supports a
healthy response and \code{unknown} when the available evidence cannot support
an attribution. Knowledge failures go to Pillar~2; the others go to the teams
that own the affected components.
The deterministic rule tree makes the initial decision
(\S\ref{sec:diagnostic}). LLMs provide evidence at fixed points in the tree,
and an agent judge audits the final route instead of replacing the tree
(\S\ref{sec:judge}).

\textbf{Pillar~4 --- Ticket resolution.} Pillar~4 assigns each ticket to one of
three resolution paths: answer it from existing knowledge, reproduce and repair
the reported problem by running a real job at \Fac{}, or route a
system-level problem to the team that owns the affected service. It then drafts
a response explaining the answer, tested repair or handoff. Support staff at
\Fac{} must review and approve the draft before it is sent to the user.

Ticket ingestion, classification and response drafting are implemented. For a
reproducible problem, ingestion also produces a validated record containing the
failing job, error signature and rerun script. The agent that consumes this
record and runs the job on a live allocation has not yet been exercised. That
path uses Apptainer~\cite{apptainer2017}, Spack~\cite{spack},
CodeAct-style~\cite{codeact2024} functions and Parsl~\cite{parsl}, and is
limited to three build attempts, thirty minutes of wall time and one node-hour.

\textbf{Pillar~2 --- Knowledge curation.} Pillar~2 turns confirmed knowledge
gaps and reusable ticket lessons into updates to the corpus used by Pillar~1.
A writer drafts each change, and a reviewer either rejects it or prepares it as
a pull request. Drafting and dry-run application are operational. Opening a
pull request is marked \stPart{} because the service account cannot merge it.
Every change therefore requires CODEOWNERS approval before it can reach the
deployed corpus. No knowledge-writer pull request has yet passed that review, so
the merge-triggered Milvus re-indexing path has not fired.

% =====================================================================

\subsection{Log Analysis and Failure Attribution}
\label{sec:diagnostic}
% =====================================================================

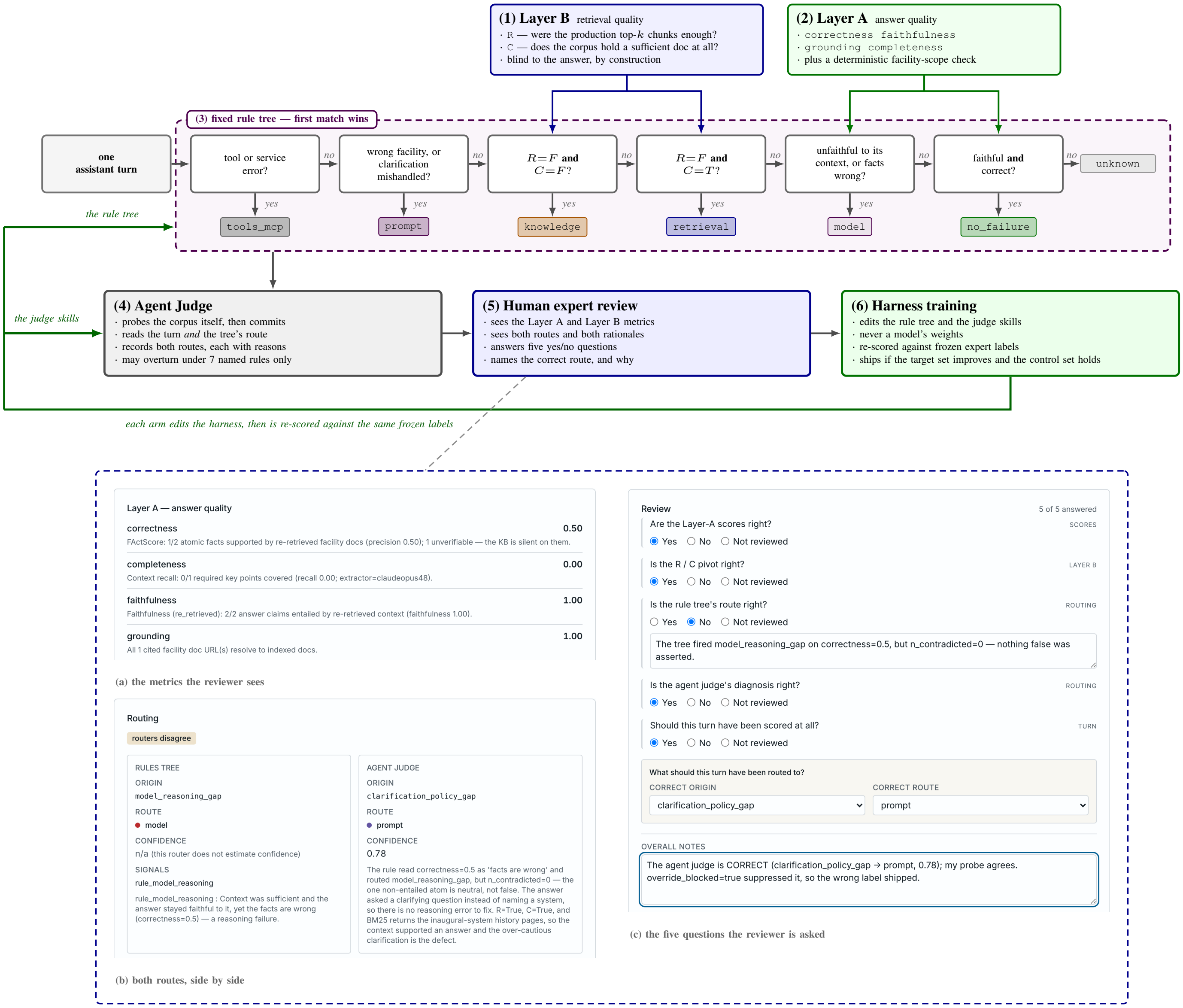
\begin{figure*}[t]
\centering
\resizebox{\textwidth}{!}{%
\begin{tikzpicture}[
  % Two type sizes only: \scriptsize\bfseries for a block title, \tiny elsewhere. One box
  % width and one pitch per row, so every arrow in that row is the same length. \bul
  % carries a \strut so inline math cannot change the bullet leading inside a box.
  font=\tiny,
  sig/.style={rounded corners=2pt, thick, align=left, inner sep=4pt,
              anchor=north, text width=4.30cm},
  layerB/.style={sig, draw=blue!55!black,  fill=blue!6},
  layerA/.style={sig, draw=green!45!black, fill=green!6},
  turn/.style={draw=black!55, rounded corners=2pt, thick, fill=black!4,
               align=center, inner sep=3pt, anchor=north,
               text width=1.95cm, minimum height=0.96cm},
  test/.style={draw=black!60, rounded corners=2pt, thick, fill=white,
               align=center, inner sep=3pt, anchor=north,
               text width=1.95cm, minimum height=0.96cm},
  route/.style={rounded corners=1pt, anchor=north, inner xsep=3pt, inner ysep=2.5pt,
                font=\tiny\ttfamily, align=center, minimum height=0.30cm},
  blk/.style={rounded corners=2pt, line width=1pt, align=left, inner sep=4.5pt,
              anchor=north west, text width=5.35cm},
  judge/.style={blk, draw=black!70, fill=black!6},
  human/.style={blk, draw=blue!55!black, fill=blue!7},
  train/.style={blk, draw=green!45!black, fill=green!8},
  ar/.style={-{Latex[length=1.6mm]}, thick, black!70},
  arb/.style={ar, draw=blue!55!black},
  arg/.style={ar, draw=green!45!black},
  arl/.style={line width=1pt, green!40!black},
  arlh/.style={arl, -{Latex[length=1.9mm]}},
  lbl/.style={font=\tiny\itshape, text=black!60},
  frame/.style={draw=violet!60!black, densely dashed, line width=0.8pt,
                rounded corners=3pt, fill=violet!4},
  tab/.style={draw=violet!60!black, line width=0.7pt, rounded corners=2pt,
              fill=white, inner xsep=4pt, inner ysep=2pt,
              font=\tiny\bfseries, text=violet!60!black}
]

% ===== (1)(2) the two signal layers =========================================
\node[layerB] (lb) at (8.750,2.05) {%
  {\scriptsize\bfseries (1) Layer B}\ \ retrieval quality\\[1pt]
  \bul \code{R} --- were the production top-$k$ chunks enough?\\
  \bul \code{C} --- does the corpus hold a sufficient doc at all?\\
  \bul blind to the answer, by construction};
\node[layerA] (la) at (13.750,2.05) {%
  {\scriptsize\bfseries (2) Layer A}\ \ answer quality\\[1pt]
  \bul \code{correctness} \ \code{faithfulness}\\
  \bul \code{grounding} \ \code{completeness}\\
  \bul plus a deterministic facility-scope check};

% ===== (3) the rule tree, as one dashed block ===============================
% Each test emits exactly one route; the order is the shipped rule order.
\node[turn] (turn) at (0.000,-0.15) {\textbf{one}\\\textbf{assistant turn}};
\node[test] (t1) at (2.500,-0.15) {tool or service\\error?};
\node[test] (t2) at (5.000,-0.15) {wrong facility, or\\clarification\\mishandled?};
\node[test] (t3) at (7.500,-0.15) {$R{=}F$ \textbf{and}\\$C{=}F$?};
\node[test] (t4) at (10.000,-0.15) {$R{=}F$ \textbf{and}\\$C{=}T$?};
\node[test] (t5) at (12.500,-0.15) {unfaithful to its\\context, or facts\\wrong?};
\node[test] (t6) at (15.000,-0.15) {faithful \textbf{and}\\correct?};
\node[route, draw=rtUnkS, fill=rtUnkF, text width=1.05cm] (r7) at (17.010,-0.48) {unknown};

\node[route, draw=rtToolsS,  fill=rtToolsF]  (r1) at (2.500,-1.54) {tools\_mcp};
\node[route, draw=rtPromptS, fill=rtPromptF] (r2) at (5.000,-1.54) {prompt};
\node[route, draw=rtKnowS,   fill=rtKnowF]   (r3) at (7.500,-1.54) {knowledge};
\node[route, draw=rtRetrS,   fill=rtRetrF]   (r4) at (10.000,-1.54) {retrieval};
\node[route, draw=rtModelS,  fill=rtModelF]  (r5) at (12.500,-1.54) {model};
\node[route, draw=rtNoFailS, fill=rtNoFailF] (r6) at (15.000,-1.54) {no\_failure};

\begin{scope}[on background layer]
  \node[frame, fit=(t1)(r7)(r1)(r6), inner sep=0.24cm] (tree) {};
\end{scope}
\node[tab, anchor=center] at ($(tree.north west)+(1.80,0)$)
  {(3) fixed rule tree --- first match wins};

\draw[ar] (turn.east) -- (t1.west);
\foreach \a/\b in {t1/t2, t2/t3, t3/t4, t4/t5, t5/t6} {
  \draw[ar] (\a.east) -- (\b.west) node[lbl, midway, above=-1pt] {no};
}
\draw[ar] (t6.east) -- (r7.west) node[lbl, midway, above=-1pt] {no};
\foreach \t/\r in {t1/r1, t2/r2, t3/r3, t4/r4, t5/r5, t6/r6} {
  \draw[ar] (\t.south) -- (\r.north) node[lbl, midway, right=1pt] {yes};
}

\draw[arb] (lb.south) -- ++(0,-0.26) -| (t3.north);
\draw[arb] (lb.south) -- ++(0,-0.26) -| (t4.north);
\draw[arg] (la.south) -- ++(0,-0.26) -| (t5.north);
\draw[arg] (la.south) -- ++(0,-0.26) -| (t6.north);

% ===== (4)(5)(6): the judge, the gate, and what the gate feeds ==============
\node[judge] (jd) at (-0.050,-2.76) {%
  {\scriptsize\bfseries (4) Agent Judge}\\[1pt]
  \bul probes the corpus itself, then commits\\
  \bul reads the turn \emph{and} the tree's route\\
  \bul records both routes, each with reasons\\
  \bul may overturn under 7 named rules only};
\node[human] (hu) at (6.150,-2.76) {%
  {\scriptsize\bfseries (5) Human expert review}\\[1pt]
  \bul sees the Layer~A and Layer~B metrics\\
  \bul sees both routes and both rationales\\
  \bul answers five yes/no questions\\
  \bul names the correct route, and why};
\node[train] (tr) at (12.350,-2.76) {%
  {\scriptsize\bfseries (6) Harness training}\\[1pt]
  \bul edits the rule tree and the judge skills\\
  \bul never a model's weights\\
  \bul re-scored against frozen expert labels\\
  \bul ships if the target set improves and the control set holds};

\draw[ar] (tree.south -| jd.north) -- (jd.north);
\draw[ar] (jd.east) -- (hu.west);
\draw[ar] (hu.east) -- (tr.west);

% The loop returns to exactly the two things an arm edits. One riser, two branches,
% each terminating on a block's left edge -- no arrowhead sits mid-line.
\coordinate (rl) at ($(tr.south)+(0,-0.55)$);
\draw[arl] (tr.south) -- (tr.south |- rl) -- (-1.72,0 |- rl);
\draw[arl] (-1.72,0 |- rl) -- (-1.72,0 |- r1);
\draw[arlh] (-1.72,0 |- jd.west) -- (jd.west);
\draw[arlh] (-1.72,0 |- r1) -- (tree.west |- r1);
\node[lbl, text=green!35!black, anchor=south west] at ($(tree.west |- r1)+(-1.62,0.04)$)
  {the rule tree};
\node[lbl, text=green!35!black, anchor=south west] at ($(jd.west)+(-1.62,0.04)$)
  {the judge skills};
\node[lbl, text=green!35!black, anchor=north west] at ($(rl)+(-15.00,-0.05)$)
  {each arm edits the harness, then is re-scored against the same frozen labels};

% ===== the zoom: the review page the human works in =========================
% Each sub-caption hangs off its OWN image's bottom-left, so it cannot drift toward
% the panel above it as image heights change.
\node[anchor=north west] (za) at (0.000,-5.98)
  {\includegraphics[width=8.10cm]{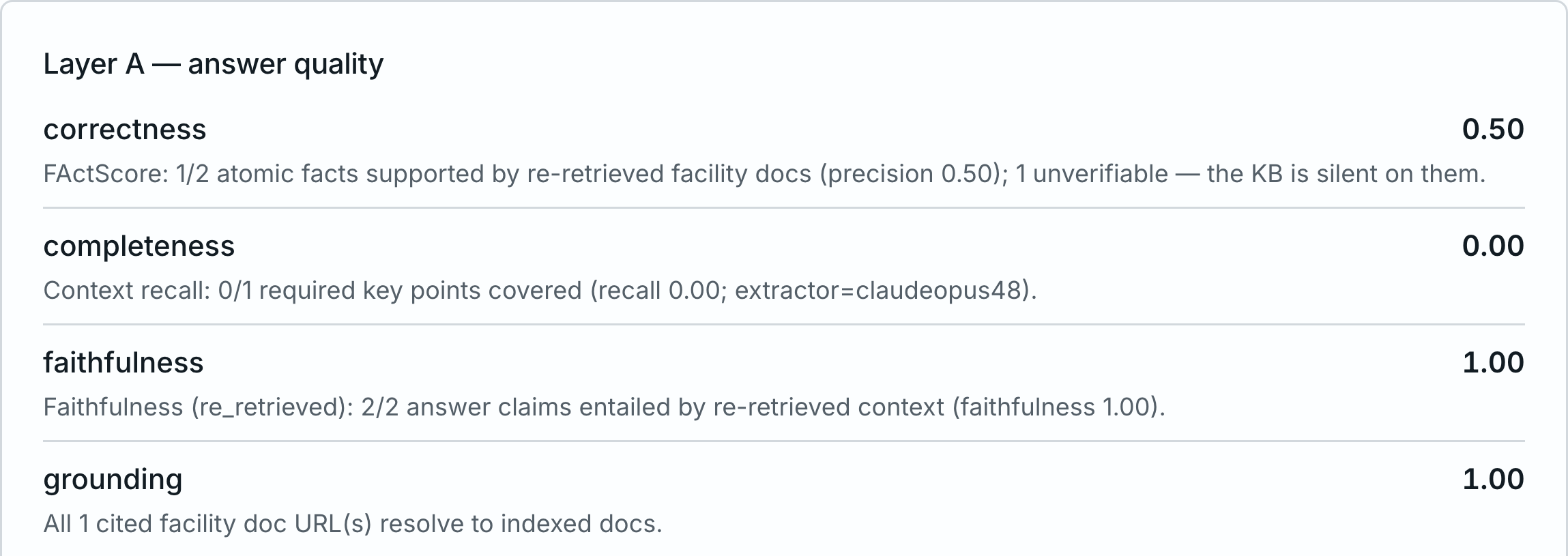}};
\node[font=\tiny\bfseries, text=black!60, anchor=north west] (ca)
  at ($(za.south west)+(0.03,-0.06)$) {(a) the metrics the reviewer sees};
\node[anchor=north west] (zb) at (0.000,-9.50)
  {\includegraphics[width=8.10cm]{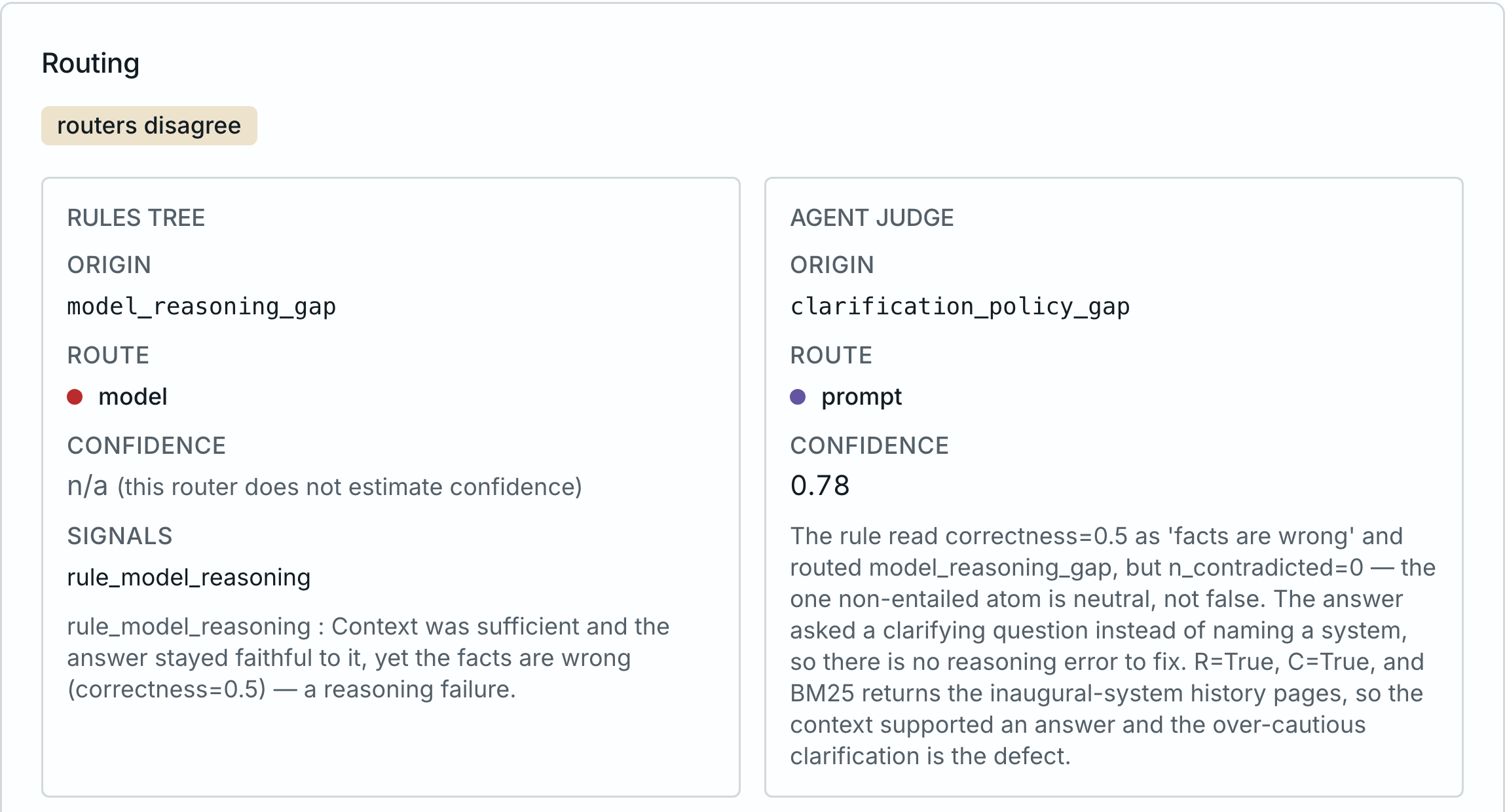}};
\node[font=\tiny\bfseries, text=black!60, anchor=north west] (cb)
  at ($(zb.south west)+(0.03,-0.06)$) {(b) both routes, side by side};
\node[anchor=north west] (zc) at (8.650,-5.98)
  {\includegraphics[width=8.10cm]{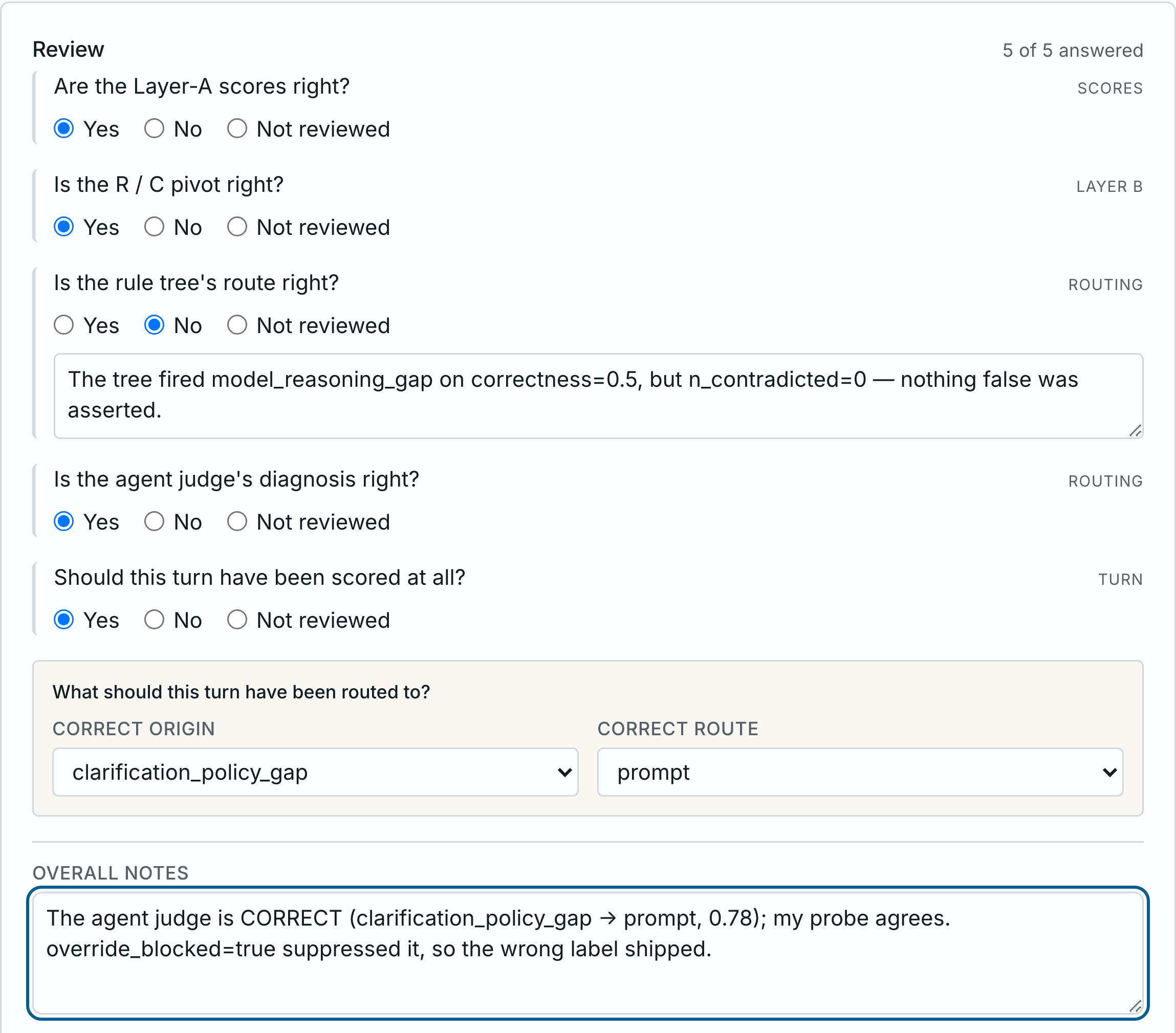}};
\node[font=\tiny\bfseries, text=black!60, anchor=north west] (cc)
  at ($(zc.south west)+(0.03,-0.06)$) {(c) the five questions the reviewer is asked};
\node[draw=blue!55!black, densely dashed, line width=0.7pt, rounded corners=2pt,
      fit=(za)(zb)(zc)(ca)(cb)(cc), inner sep=5pt] (zoombox) {};
\draw[densely dashed, black!45, line width=0.6pt]
  ($(hu.south west)+(0.9,0)$) -- ($(zoombox.north west)+(5.55,0)$);
\end{tikzpicture}%
}
\caption{The diagnostic workflow and its expert-review loop. Layers~B and A
produce retrieval and answer evidence, respectively. The ordered rule tree (3)
assigns the initial route, and the agent judge (4) audits that decision. A human
expert reviews the evidence and both rationales (5); those labels support
harness changes to the rule tree and judge skills (6). The lower panels show the
review interface for a turn on which the two routes disagree.}
\label{fig:pipeline}
\end{figure*}

Pillar~3 converts each logged interaction into a diagnosis and an actionable
route. Figure~\ref{fig:pipeline} summarizes the workflow. Layer~A evaluates the
answer, while Layer~B evaluates retrieval without seeing the answer. An ordered
rule tree combines this evidence and assigns the turn to a component owner. An
agent judge audits that decision, and a human expert reviews disputed or
uncertain cases. The resulting expert labels are then used to improve the rule
tree and the judge skills.

\subsubsection{Answer-Quality Evidence (Layer A)}
\label{sec:layerA}

Layer~A computes four answer-quality signals and one deterministic scope check.
All entailment decisions use the same three-class judge
(entailment, neutral or contradiction), which keeps differences between signals
attributable to their evidence rather than to different judging functions.

\code{correctness} measures the fraction of answer claims supported by evidence
retrieved separately for each claim. \code{faithfulness} applies the same test
to the context originally shown to the model. Their difference is therefore
diagnostic: an answer may contain correct information that was not supported by
its supplied context.

\code{grounding} checks whether paths, modules and documentation URLs cited in
the answer exist. Following the fabricated-citation setting of
ALCE~\cite{alce2023}, it returns \emph{found}, \emph{mangled\_real} or
\emph{fabricated}. The middle value indicates that the artifact exists but was
damaged during chunking, which is a retrieval defect rather than a fabricated
citation. An answer with no such citation receives \code{n/a}, not zero.
\code{completeness} follows the recall formulation of
RAGChecker~\cite{ragchecker2024}. Finally, \code{facility\_scope} is a
deterministic lexicon check that rejects answers referring to another facility
or to an unsupported GPU stack.

The abstention detector distinguishes \emph{full}, \emph{partial} and
\emph{no} abstention. It removes fenced code, references and clarifying
questions before counting substantive claims. Only a full abstention exempts
the answer-quality signals. Ambiguous cases are treated as partial abstentions,
because incorrectly exempting a substantive answer could hide a genuine
failure.

\subsubsection{Retrieval and Corpus Sufficiency (Layer B)}
\label{sec:layerB}

Layer~B separates two questions that are often combined. $R$ asks whether the
top-$k$ passages returned in production were sufficient. $C$ asks whether the
required information existed anywhere in the corpus. $R$ follows the
Sufficient Context test~\cite{sufficientcontext2024}; $C$ extends that test to
a wider candidate set obtained from expanded dense queries and BM25 search.

The two verdicts determine the responsible component. If $R$ is true, retrieval
was sufficient and the rule tree evaluates the generation-side evidence. If
$R$ is false but $C$ is true, the corpus contains the answer and the failure is
assigned to \code{retrieval}. If both are false, the corpus lacks sufficient
documentation and the failure is assigned to \code{knowledge}. This distinction
turns a sufficiency judgment into an actionable work item.

Two constraints preserve that interpretation. First, Layer~B never sees the
generated answer; otherwise it could not distinguish insufficient retrieval
from poor answer generation. Second, $R$ and $C$ use different query policies.
$R$ evaluates the query and passages used in production. $C$ may use rewritten,
goal-directed queries because it tests corpus coverage rather than production
retrieval. A relaxed condition is therefore applied to $R$ only when supported
by direct production evidence.

For complex questions, $C$ is evaluated over answer requirements rather than by
asking whether a large collection of passages is sufficient as a whole. The
system decomposes the question into requirements, marks the essential ones as
\emph{core}, and checks each core requirement against the corpus. Core status is
assigned without access to coverage results, preventing a missing fact from
being reclassified as unimportant. If requirement extraction is malformed, all
requirements are treated as core, so the system fails conservatively.

We use semantic core requirements rather than a fixed fractional coverage
threshold, which would treat requirements of unequal importance as
interchangeable. The materiality judgment preserves the meaning of corpus
sufficiency across questions with different structures.

\subsubsection{Cause Attribution with an Ordered Rule Tree}
\label{sec:ruletree}

The router is an ordered list of 28 rules with first-match semantics. A predicate
over a missing signal evaluates to false, so absent evidence cannot trigger a
route. The ordering in Fig.~\ref{fig:pipeline} reflects four priorities. Service
and tool failures are checked first, because a backend outage should not be
evaluated as a model-generated answer. Deterministic facility-scope and
clarification checks come next. The $R\times C$ sufficiency pivot then separates
retrieval and knowledge failures. Answer-quality signals are considered only
after retrieval has been shown to be sufficient.

This ordering prevents answer-quality signals from absorbing retrieval errors.
When retrieval is insufficient, correctness and completeness are also likely to
be low. Evaluating those signals before the sufficiency pivot would therefore
assign retrieval failures to generation-side components. Sufficient retrieval
is treated as a prerequisite for attributing a failure to answer generation.

On the generation side, the tree distinguishes how the answer used its context
from what the answer asserted. A fabricated citation or a contradiction of the
provided context is assigned to \code{model}, because the required evidence was
available but not used correctly. An answer that remains faithful to its context
but reaches an incorrect conclusion is also assigned to \code{model}, for a
different underlying reason. No answer-quality score directly produces a
\code{prompt} route, because a low score alone cannot distinguish an inadequate
instruction from inadequate execution. Prompt routes arise from explicit policy
checks performed earlier in the tree.

The terminal outcomes \code{no\_failure} and \code{unknown} have different
meanings. \code{no\_failure} requires positive evidence that retrieval was
sufficient and the answer was faithful and correct. \code{unknown} indicates
that no supported attribution could be made. Keeping these outcomes separate
distinguishes a verified healthy turn from one that could not be evaluated. If
a deterministic signal conflicts with an unreliable judged signal, the tree
escalates the turn for human review rather than guessing.

Table~\ref{tab:routes} states the seven outcomes the tree can emit. Five are
failure modes, each naming the component that must change; two are terminal and
dispatch nobody. Every turn receives exactly one, so no turn leaves the pipeline
without a recorded disposition.

\begin{table*}[t]
\caption{The seven attribution outcomes the rule tree can emit, and the evidence
that decides each. $R$ is retrieval sufficiency at the production result and $C$
is sufficiency over the whole corpus (\S\ref{sec:layerB}); T and F are their
truth values. Every scored turn receives exactly one outcome.}
\label{tab:routes}
\centering
\footnotesize
\setlength{\tabcolsep}{5pt}
\begin{tabular}{@{}l p{12.6cm} l@{}}
\toprule
Outcome & What it asserts, and the evidence that decides it & Repaired by \\
\midrule
\multicolumn{3}{@{}l}{\textbf{Failure modes} --- the turn failed, and the outcome names the component that can repair it} \\[2pt]
\code{knowledge} & No authoritative document exists anywhere in the corpus: the
production result is insufficient and a wide corpus probe fails to surface one
either ($R{=}$F, $C{=}$F). The fix is to write the missing page. & documentation \\[2pt]
\code{retrieval} & A sufficient document does exist, but production retrieval did
not return it ($R{=}$F, $C{=}$T). Internal conditions separate ranking depth,
indexing, and query formulation, which are three different fix sites. & retrieval \\[2pt]
\code{prompt} & A deterministic answer-policy check fires ahead of the
sufficiency pivot: the answer addresses the wrong facility or the wrong system,
or a critical user detail was missing and no clarifying question was
asked. & prompt policy \\[2pt]
\code{model} & Retrieval was sufficient and generation still failed: the answer
cites a path or module that exists nowhere in the corpus, contradicts its own
supplied context, or stays faithful to that context yet reaches a wrong
conclusion. & model serving \\[2pt]
\code{tools\_mcp} & A tool call or backend service failed --- error, schema
mismatch, permission, or outage --- so the turn never had a fair chance to
succeed. Checked first, so an outage is never scored as a bad answer. & platform \\
\midrule
\multicolumn{3}{@{}l}{\textbf{Terminal outcomes} --- the turn is concluded and no work is dispatched} \\[2pt]
\code{no\_failure} & Positively verified healthy: retrieval was sufficient and the
answer was faithful and correct. This is an affirmative finding, not the absence
of one. & --- \\[2pt]
\code{unknown} & No attribution the evidence supports. The tree's fall-through
lands here, so the turn is reported as unattributed rather than assigned to a
team on a guess. & --- (escalated) \\
\bottomrule
\end{tabular}
\end{table*}

Each top-level route retains more specific internal conditions for engineering
remediation. The paper reports the five component-level causes for clarity. A
cause is included only when a corresponding detector can provide supporting
evidence.

\subsubsection{Agentic Audit}
\label{sec:judge}

After the rule tree assigns a route, an \emph{agent judge} audits the decision.
The judge does not replace the router. It checks whether the evidence used by
the matched rule supports the assigned cause. When it disagrees, it identifies
the disputed signal, producing a concrete item for review rather than only a
second route label.

We selected this audit design instead of a second generative router because a
second router would consume the same potentially unreliable signals and would
not provide an independent check on them.

The agent judge differs from the single-shot judges used to compute individual
signals. It runs a bounded probe loop over the corpus: it lists candidate URLs,
searches for evidence and reads selected chunks. It can therefore find evidence
that was absent from the production retrieval result. As evaluated in
\S\ref{sec:overturn}, this access is most useful when the rule tree reached a
decision from missing evidence.

Every judge call is logged, observations are size-limited, and probe results are
content-addressed and pinned for replay (\S\ref{sec:repro}). These controls make
each conclusion traceable to evidence the agent actually inspected.

Authority is configured separately from judgment. The agent judge may overturn
the tree only for seven rule classes whose inputs it can independently verify.
For all other classes, its disagreement is recorded for human review without
changing the route. The gate depends on the rule that fired, not on the judge's
self-reported confidence. Section~\ref{sec:overturn} evaluates this policy.

\subsubsection{Reproducibility Controls}
\label{sec:repro}

The corpus probe used to compute $C$ expands each question into sub-queries with
an LLM call requesting \code{temperature=0}. The inference gateway did not
honor that setting: two otherwise identical runs produced different expansion
sets for all 78 tested questions, and the resulting retrieval differences
changed the final route for 16\% of turns.

We therefore cache each expansion under a hash of the query, model and prompt.
The cache is committed with the evaluation artifacts, making expansions
identical across runs and machines and reviewable as a diff. Every result in
this paper uses these pinned expansions.

% =====================================================================

\section{Human-Guided Self-Evolution}
\label{sec:control}
% =====================================================================

Production evidence does not change the system automatically. Instead, experts
review diagnoses, record corrections and propose bounded changes to the rules,
prompts or knowledge base. Each proposal is evaluated against a fixed reference
before it is accepted. Figure~\ref{fig:harness} summarizes this cycle. We call
the process \emph{governed self-evolution} because the system learns from its
operation while people retain authority over every consequential change.

\subsection{Expert Review and Reference Labels}

A diagnostic route is a proposal rather than an action. The review service
therefore presents the evidence behind each route before a human decision is
recorded. It provides two role-scoped workspaces behind \FacAdj{} single sign-on:
an administrative workspace for reviewing chatbot diagnoses and a staff
workspace for reviewing ticket responses. Both read immutable analysis
artifacts and write decisions to a separate append-only store.

For each chatbot turn, the administrative workspace shows the answer-quality
signals, the retrieval and corpus-sufficiency verdicts, the matched rule, and
the agent judge's independent assessment. Reviewers can accept the route,
provide a corrected route and explanation, or mark the sample as unsuitable for
evaluation. An unanswered item remains \emph{uncertain}; it is never interpreted
as agreement. These decisions form the expert reference used for harness
training.

\subsection{Ticket Feedback and Knowledge Accumulation}

The staff workspace applies the same approval principle to support tickets. It
re-triages each ticket, retrieves supporting documentation and drafts a response
with citations derived from the retrieved sources. A support expert reviews the
draft before exporting it to the expert's mail client; the system does not send
messages automatically. Reusable lessons and confirmed documentation gaps are
then passed to the Pillar~2 writer--reviewer workflow. Any resulting corpus
change requires CODEOWNERS approval before merge and re-indexing.

\subsection{Harness Training}
\label{sec:harness}
\label{sec:improvementloop}
% =====================================================================

Harness training improves the diagnostic pipeline without changing model
weights. The editable harness consists of the rule tree, the judge skills and
the override gate. A human reviewer identifies a recurring error, proposes one
bounded change and names the turns that should improve and the control turns
that should remain unchanged. One complete proposal-and-evaluation cycle is an
\emph{arm}.

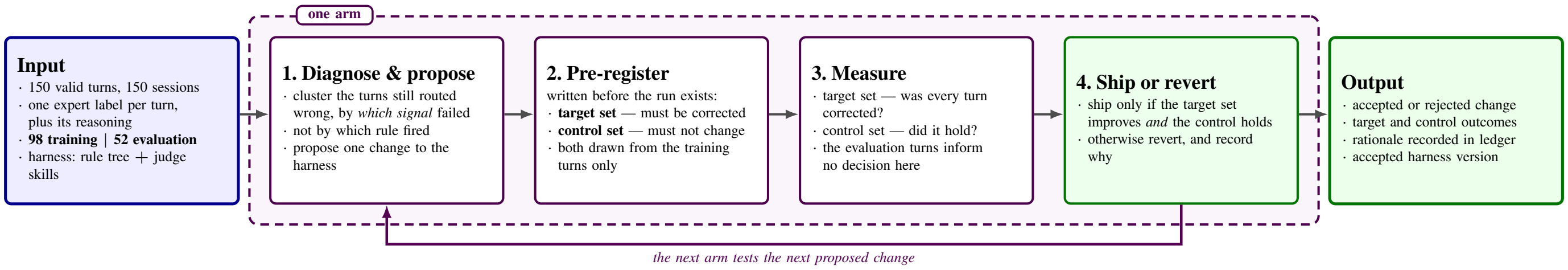
\begin{figure*}[t]
\centering
\resizebox{\textwidth}{!}{%
\begin{tikzpicture}[
  % Same conventions as Fig.~\ref{fig:pipeline}: two type sizes, one box width and one
  % pitch across the row so every arrow is the same length, \bul for the leading.
  font=\tiny,
  cell/.style={rounded corners=2pt, align=left, inner sep=4pt,
               anchor=north, text width=2.55cm, minimum height=2.02cm},
  io/.style={cell, draw=blue!55!black, line width=1pt, fill=blue!7},
  stn/.style={cell, draw=violet!60!black, thick, fill=white},
  ship/.style={cell, draw=green!45!black, thick, fill=green!7},
  outbox/.style={cell, draw=green!45!black, line width=1pt, fill=green!8},
  ar/.style={-{Latex[length=1.6mm]}, thick, black!70},
  arv/.style={-{Latex[length=1.6mm]}, line width=0.9pt, violet!65!black},
  lbl/.style={font=\tiny\itshape, text=black!60},
  frame/.style={draw=violet!60!black, densely dashed, line width=0.8pt,
                rounded corners=3pt, fill=violet!4},
  tab/.style={draw=violet!60!black, line width=0.7pt, rounded corners=2pt,
              fill=white, inner xsep=4pt, inner ysep=2pt,
              font=\tiny\bfseries, text=violet!60!black}
]
\node[io]     (in) at (0.00,0.00) {%
  {\scriptsize\bfseries Input}\\[1pt]
  \bul 150 valid turns, 150 sessions\\
  \bul one expert label per turn,\\ \phantom{\bul}plus its reasoning\\
  \bul \textbf{98 training} $\mid$ \textbf{52 evaluation}\\
  \bul harness: rule tree $+$ judge\\ \phantom{\bul}skills};
\node[stn]    (s1) at (3.22,0.00) {%
  {\scriptsize\bfseries 1. Diagnose \& propose}\\[1pt]
  \bul cluster the turns still routed\\ \phantom{\bul}wrong, by \emph{which signal} failed\\
  \bul not by which rule fired\\
  \bul propose one change to the\\ \phantom{\bul}harness};
\node[stn]    (s2) at (6.44,0.00) {%
  {\scriptsize\bfseries 2. Pre-register}\\[1pt]
  written before the run exists:\\
  \bul \textbf{target set} --- must be corrected\\
  \bul \textbf{control set} --- must not change\\
  \bul both drawn from the training\\ \phantom{\bul}turns only};
\node[stn]    (s3) at (9.66,0.00) {%
  {\scriptsize\bfseries 3. Measure}\\[1pt]
  \bul target set --- was every turn\\ \phantom{\bul}corrected?\\
  \bul control set --- did it hold?\\
  \bul the evaluation turns inform\\ \phantom{\bul}no decision here};
\node[ship]   (s4) at (12.88,0.00) {%
  {\scriptsize\bfseries 4. Ship or revert}\\[1pt]
  \bul ship only if the target set\\ \phantom{\bul}improves \emph{and} the control holds\\
  \bul otherwise revert, and record\\ \phantom{\bul}why};
\node[outbox] (ob) at (16.10,0.00) {%
  {\scriptsize\bfseries Output}\\[1pt]
  \bul accepted or rejected change\\
  \bul target and control outcomes\\
  \bul rationale recorded in ledger\\
  \bul accepted harness version};

\begin{scope}[on background layer]
  \node[frame, fit=(s1)(s4), inner sep=0.24cm] (arm) {};
\end{scope}
\node[tab, anchor=center] at ($(arm.north west)+(1.05,0)$) {one arm};

% Arrows are pinned to a fixed offset from the row's top edge, not to each box's
% vertical middle, so an unequal box height cannot tilt them out of line.
\foreach \a/\b in {in/s1, s1/s2, s2/s3, s3/s4, s4/ob} {
  \draw[ar] ($(\a.north east)+(0,-0.95)$) -- ($(\b.north west)+(0,-0.95)$);
}
\draw[arv] (s4.south) -- ++(0,-0.48) -| ($(s1.south)+(0,-0.48)$) -- (s1.south);
\node[lbl, text=violet!60!black, anchor=north] at (8.05,-2.50)
  {the next arm tests the next proposed change};
\end{tikzpicture}%
}
\caption{Human-guided harness training. Each proposed change is evaluated as
one arm with pre-registered target and control sets. A change is accepted only
when the target cases improve and the controls remain stable. Evaluation cases
are not used to propose, tune or select an arm.}
\label{fig:harness}
\end{figure*}

As Fig.~\ref{fig:harness} shows, each arm follows four steps: diagnose an error,
pre-register its target and control sets, run the revised harness, and either
accept or reject the change. Evaluation turns remain hidden from design
decisions. This process borrows the train/evaluation split and experiment ledger
of model training, but the trainable object is the surrounding software and
prompts rather than a model's parameters.

\subsubsection{Expert Reference and Data Split}
\label{sec:harness-data}

We sampled 150 production sessions that did not overlap the development pilot
of \S\ref{sec:pilot}. Three domain scientists labelled the turns through the
review interface, and a fourth resolved disagreements. The router scores 171
turns across these sessions; after excluding unsuitable records, 150 of them
carry an expert label and form the graded reference, split by session into 98
training and 52 evaluation turns. Session-level splitting prevents turns from the same conversation from
appearing on both sides.

The reference is generated from the reviewer records, content-hashed and checked
for drift before every evaluation. Reviewers provide both a route and a written
explanation. The explanation is essential: a corrected label identifies the
error, while the accompanying rationale suggests what rule, instruction or
evidence source should change. Section~\ref{sec:eval} reports the resulting
campaign and its accepted and rejected arms.

\section{Evaluation}
\label{sec:eval}
% =====================================================================

We evaluate four questions: whether log analysis produces actionable diagnoses,
whether human-guided harness training improves those diagnoses, what the agent
judge contributes, and whether the resulting diagnoses lead to effective
repairs. A 97-turn pilot supported taxonomy development but not claims of
generalization. Formal results use a separate 150-turn multi-expert reference
with a session-disjoint training/evaluation split, together with the full corpus
of 2{,}174 scored turns.

\subsection{Evaluation Data and Development Pilot}
\label{sec:pilot}

The router was first developed against 100 sessions collected between February
and July 2026, with one scored turn per session and stratification across the
five failure modes and \code{no\_failure}. A single author reviewed four
questions per turn. Three turns were excluded as unscoreable, leaving 97. Initial
agreement was \textbf{53 of 97 (55\%)}, including ten false \code{knowledge}
routes. Repairing the input signals increased agreement to \textbf{77 of 97
(79\%)} and reduced false \code{knowledge} routes to four.

This result is developmental rather than confirmatory. The reviewer was a
single author, and the same turns informed the taxonomy, rule ordering, and
thresholds. We therefore use the separate 150-turn, multi-expert reference
described in \S\ref{sec:harness} for formal evaluation.

\subsection{RQ1: Does Log Analysis Produce Actionable Diagnoses?}
\label{sec:distribution}

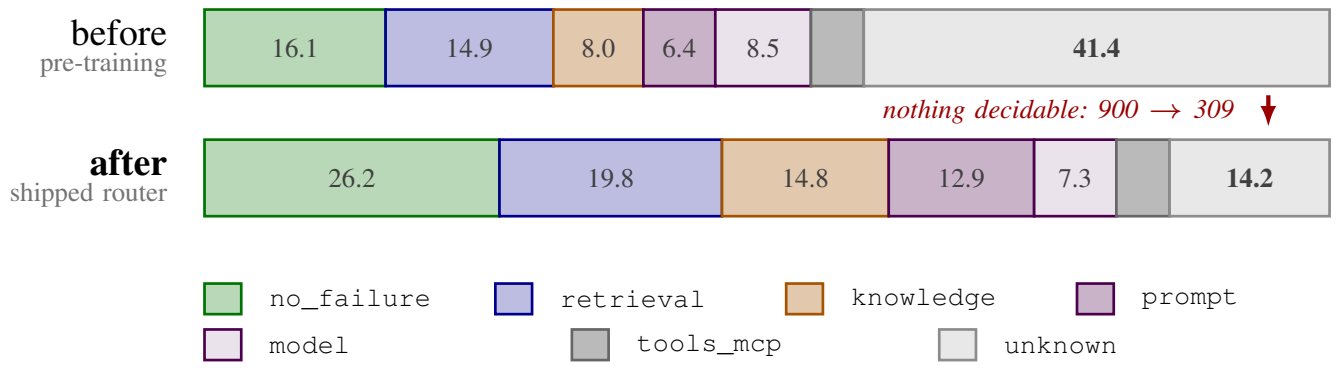
\begin{figure}[t]
\centering
\resizebox{\columnwidth}{!}{%
\begin{tikzpicture}[x=1cm,y=1cm]
\node[anchor=east, font=\scriptsize, align=right] at (-0.12,1.04) {before};
\node[anchor=east, font=\tiny, text=black!55] at (-0.12,0.83) {pre-training};
\filldraw[fill=rtNoFailF, draw=rtNoFailS, line width=0.5pt] (0.000,0.69) rectangle (1.183,1.19);
\node[font=\tiny, text=black!75] at (0.592,0.94) {16.1};
\filldraw[fill=rtRetrF, draw=rtRetrS, line width=0.5pt] (1.183,0.69) rectangle (2.279,1.19);
\node[font=\tiny, text=black!75] at (1.731,0.94) {14.9};
\filldraw[fill=rtKnowF, draw=rtKnowS, line width=0.5pt] (2.279,0.69) rectangle (2.867,1.19);
\node[font=\tiny, text=black!75] at (2.573,0.94) {8.0};
\filldraw[fill=rtPromptF, draw=rtPromptS, line width=0.5pt] (2.867,0.69) rectangle (3.337,1.19);
\node[font=\tiny, text=black!75] at (3.102,0.94) {6.4};
\filldraw[fill=rtModelF, draw=rtModelS, line width=0.5pt] (3.337,0.69) rectangle (3.962,1.19);
\node[font=\tiny, text=black!75] at (3.650,0.94) {8.5};
\filldraw[fill=rtToolsF, draw=rtToolsS, line width=0.5pt] (3.962,0.69) rectangle (4.307,1.19);
\filldraw[fill=rtUnkF, draw=rtUnkS, line width=0.5pt] (4.307,0.69) rectangle (7.350,1.19);
\node[font=\tiny, text=black!75] at (5.829,0.94) {\textbf{41.4}};
\node[anchor=east, font=\scriptsize, align=right] at (-0.12,0.19) {\textbf{after}};
\node[anchor=east, font=\tiny, text=black!55] at (-0.12,-0.02) {shipped router};
\filldraw[fill=rtNoFailF, draw=rtNoFailS, line width=0.5pt] (0.000,-0.16) rectangle (1.927,0.34);
\node[font=\tiny, text=black!75] at (0.964,0.09) {26.2};
\filldraw[fill=rtRetrF, draw=rtRetrS, line width=0.5pt] (1.927,-0.16) rectangle (3.381,0.34);
\node[font=\tiny, text=black!75] at (2.654,0.09) {19.8};
\filldraw[fill=rtKnowF, draw=rtKnowS, line width=0.5pt] (3.381,-0.16) rectangle (4.470,0.34);
\node[font=\tiny, text=black!75] at (3.925,0.09) {14.8};
\filldraw[fill=rtPromptF, draw=rtPromptS, line width=0.5pt] (4.470,-0.16) rectangle (5.420,0.34);
\node[font=\tiny, text=black!75] at (4.945,0.09) {12.9};
\filldraw[fill=rtModelF, draw=rtModelS, line width=0.5pt] (5.420,-0.16) rectangle (5.957,0.34);
\node[font=\tiny, text=black!75] at (5.688,0.09) {7.3};
\filldraw[fill=rtToolsF, draw=rtToolsS, line width=0.5pt] (5.957,-0.16) rectangle (6.305,0.34);
\filldraw[fill=rtUnkF, draw=rtUnkS, line width=0.5pt] (6.305,-0.16) rectangle (7.350,0.34);
\node[font=\tiny, text=black!75] at (6.828,0.09) {\textbf{14.2}};
\draw[-{Latex[length=1.4mm]}, thick, red!60!black] (6.95,0.63) -- (6.95,0.42);
\node[anchor=east, font=\tiny\itshape, text=red!60!black] at (6.85,0.52)
  {nothing decidable: 900 $\rightarrow$ 309};
\filldraw[fill=rtNoFailF, draw=rtNoFailS, line width=0.5pt] (0.00,-0.80) rectangle (0.24,-0.60);
\node[anchor=west, font=\tiny] at (0.30,-0.70) {\code{no\_failure}};
\filldraw[fill=rtRetrF, draw=rtRetrS, line width=0.5pt] (1.90,-0.80) rectangle (2.14,-0.60);
\node[anchor=west, font=\tiny] at (2.20,-0.70) {\code{retrieval}};
\filldraw[fill=rtKnowF, draw=rtKnowS, line width=0.5pt] (3.80,-0.80) rectangle (4.04,-0.60);
\node[anchor=west, font=\tiny] at (4.10,-0.70) {\code{knowledge}};
\filldraw[fill=rtPromptF, draw=rtPromptS, line width=0.5pt] (5.70,-0.80) rectangle (5.94,-0.60);
\node[anchor=west, font=\tiny] at (6.00,-0.70) {\code{prompt}};
\filldraw[fill=rtModelF, draw=rtModelS, line width=0.5pt] (0.00,-1.10) rectangle (0.24,-0.90);
\node[anchor=west, font=\tiny] at (0.30,-1.00) {\code{model}};
\filldraw[fill=rtToolsF, draw=rtToolsS, line width=0.5pt] (2.40,-1.10) rectangle (2.64,-0.90);
\node[anchor=west, font=\tiny] at (2.70,-1.00) {\code{tools\_mcp}};
\filldraw[fill=rtUnkF, draw=rtUnkS, line width=0.5pt] (4.80,-1.10) rectangle (5.04,-0.90);
\node[anchor=west, font=\tiny] at (5.10,-1.00) {\code{unknown}};
\end{tikzpicture}%
}
\caption{The same 2{,}174 turns scored before and after harness training, on identical
input. Turns the router could not attribute fall from \textbf{900 (41.4\%) to 309 (14.2\%)}
because the retrieval test now returns a verdict where it previously returned nothing.}
\label{fig:corpusdist}
\end{figure}

We applied the shipped router to the complete interaction corpus of
\textbf{1{,}229 sessions and 2{,}174 scored assistant turns}. Because the same
turns had been scored before harness training, Fig.~\ref{fig:corpusdist}
provides a controlled comparison on identical inputs.

The principal improvement is a reduction in unattributed turns. The tree could
not assign a failure mode to \textbf{900 turns (41.4\%)} before training and
\textbf{309 turns (14.2\%)} afterward; attributed turns therefore increased
from 924 to \textbf{1{,}295}. Run artifacts identify the mechanism: 729 turns
in the earlier run had no retrieval verdict, and 684 of them consequently fell
through to \code{unknown}. With the revised retrieval test, their largest
destinations are
\code{no\_failure} (198), \code{knowledge} (150), \code{retrieval} (147) and \code{prompt}
(75), while 106 remain unattributed. Thus, the change reflects improved evidence
coverage rather than a more permissive attribution policy.

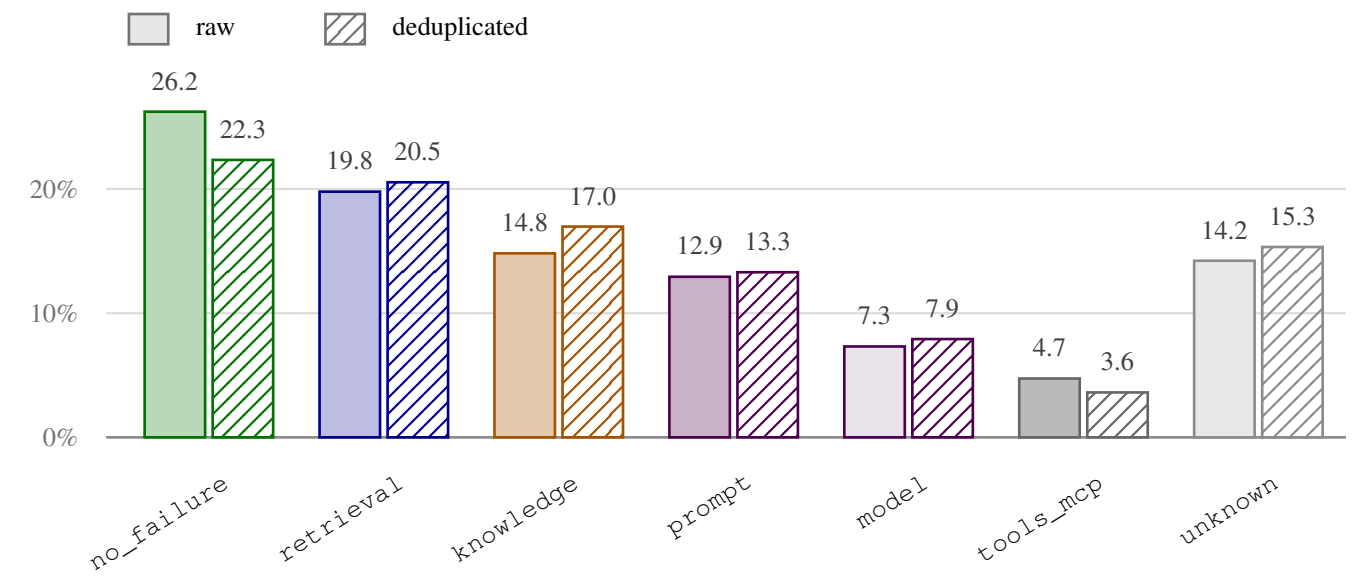
\begin{figure}[t]
\centering
% Canvas is sized to \columnwidth on purpose: no \resizebox, so \tiny renders at
% 5pt rather than being scaled down into illegibility.
\begin{tikzpicture}[x=1cm,y=1cm]
\draw[black!14, line width=0.4pt] (0.52,0.000) -- (8.72,0.000);
\node[anchor=east, font=\tiny, text=black!55] at (0.46,0.000) {0\%};
\draw[black!14, line width=0.4pt] (0.52,0.821) -- (8.72,0.821);
\node[anchor=east, font=\tiny, text=black!55] at (0.46,0.821) {10\%};
\draw[black!14, line width=0.4pt] (0.52,1.643) -- (8.72,1.643);
\node[anchor=east, font=\tiny, text=black!55] at (0.46,1.643) {20\%};
\draw[black!45, line width=0.5pt] (0.52,0) -- (8.72,0);
\filldraw[fill=rtNoFailF, draw=rtNoFailS, line width=0.5pt] (0.774,0) rectangle (1.174,2.154);
\filldraw[fill=rtNoFailF, draw=rtNoFailS, line width=0.5pt, pattern=north east lines, pattern color=rtNoFailS] (1.224,0) rectangle (1.624,1.835);
\node[font=\tiny, text=black!75, anchor=south] at (0.974,2.174) {26.2};
\node[font=\tiny, text=black!75, anchor=south] at (1.424,1.855) {22.3};
\node[font=\tiny\ttfamily, rotate=32, anchor=north east] at (1.319,-0.06) {no\_failure};
\filldraw[fill=rtRetrF, draw=rtRetrS, line width=0.5pt] (1.931,0) rectangle (2.331,1.625);
\filldraw[fill=rtRetrF, draw=rtRetrS, line width=0.5pt, pattern=north east lines, pattern color=rtRetrS] (2.381,0) rectangle (2.781,1.687);
\node[font=\tiny, text=black!75, anchor=south] at (2.131,1.645) {19.8};
\node[font=\tiny, text=black!75, anchor=south] at (2.581,1.707) {20.5};
\node[font=\tiny\ttfamily, rotate=32, anchor=north east] at (2.476,-0.06) {retrieval};
\filldraw[fill=rtKnowF, draw=rtKnowS, line width=0.5pt] (3.088,0) rectangle (3.488,1.217);
\filldraw[fill=rtKnowF, draw=rtKnowS, line width=0.5pt, pattern=north east lines, pattern color=rtKnowS] (3.538,0) rectangle (3.938,1.394);
\node[font=\tiny, text=black!75, anchor=south] at (3.288,1.237) {14.8};
\node[font=\tiny, text=black!75, anchor=south] at (3.738,1.414) {17.0};
\node[font=\tiny\ttfamily, rotate=32, anchor=north east] at (3.633,-0.06) {knowledge};
\filldraw[fill=rtPromptF, draw=rtPromptS, line width=0.5pt] (4.245,0) rectangle (4.645,1.062);
\filldraw[fill=rtPromptF, draw=rtPromptS, line width=0.5pt, pattern=north east lines, pattern color=rtPromptS] (4.695,0) rectangle (5.095,1.092);
\node[font=\tiny, text=black!75, anchor=south] at (4.445,1.082) {12.9};
\node[font=\tiny, text=black!75, anchor=south] at (4.895,1.112) {13.3};
\node[font=\tiny\ttfamily, rotate=32, anchor=north east] at (4.790,-0.06) {prompt};
\filldraw[fill=rtModelF, draw=rtModelS, line width=0.5pt] (5.402,0) rectangle (5.802,0.601);
\filldraw[fill=rtModelF, draw=rtModelS, line width=0.5pt, pattern=north east lines, pattern color=rtModelS] (5.852,0) rectangle (6.252,0.650);
\node[font=\tiny, text=black!75, anchor=south] at (5.602,0.621) {7.3};
\node[font=\tiny, text=black!75, anchor=south] at (6.052,0.670) {7.9};
\node[font=\tiny\ttfamily, rotate=32, anchor=north east] at (5.947,-0.06) {model};
\filldraw[fill=rtToolsF, draw=rtToolsS, line width=0.5pt] (6.559,0) rectangle (6.959,0.389);
\filldraw[fill=rtToolsF, draw=rtToolsS, line width=0.5pt, pattern=north east lines, pattern color=rtToolsS] (7.009,0) rectangle (7.409,0.297);
\node[font=\tiny, text=black!75, anchor=south] at (6.759,0.409) {4.7};
\node[font=\tiny, text=black!75, anchor=south] at (7.209,0.317) {3.6};
\node[font=\tiny\ttfamily, rotate=32, anchor=north east] at (7.104,-0.06) {tools\_mcp};
\filldraw[fill=rtUnkF, draw=rtUnkS, line width=0.5pt] (7.716,0) rectangle (8.116,1.168);
\filldraw[fill=rtUnkF, draw=rtUnkS, line width=0.5pt, pattern=north east lines, pattern color=rtUnkS] (8.166,0) rectangle (8.566,1.259);
\node[font=\tiny, text=black!75, anchor=south] at (7.916,1.188) {14.2};
\node[font=\tiny, text=black!75, anchor=south] at (8.366,1.279) {15.3};
\node[font=\tiny\ttfamily, rotate=32, anchor=north east] at (8.261,-0.06) {unknown};
\filldraw[fill=black!10, draw=black!55, line width=0.5pt] (0.67,2.60) rectangle (0.93,2.80);
\node[anchor=west, font=\tiny] at (0.99,2.70) {raw};
\filldraw[fill=black!10, draw=black!55, line width=0.5pt, pattern=north east lines, pattern color=black!55] (1.97,2.60) rectangle (2.23,2.80);
\node[anchor=west, font=\tiny] at (2.29,2.70) {deduplicated};
\end{tikzpicture}
\caption{Failure-mode distribution under two counting rules, as a share of each total.
\emph{Raw} is every scored turn (2{,}174); \emph{deduplicated} keeps one session per distinct
opening question (1{,}768), collapsing the 33-fold repeats so one defect is counted once. The
ordering barely changes; the shares do. \code{knowledge} is the class a raw count
understates, because documentation gaps repeat less often than live-state questions.}
\label{fig:dedup}
\end{figure}

Raw turn counts can overstate recurring problems. The 1{,}229 sessions contain
927 distinct opening questions, and the most frequent question appears 33
times. Figure~\ref{fig:dedup} therefore also reports a deduplicated view. In
total, 645 question groups contain at least one non-healthy route; the five
largest groups account for 106 sessions and concern job and node counts,
outbound network access, login procedures, and job submission. This grouping
turns per-turn diagnoses into a support work queue without allowing repeated
questions to dominate it.

Two caveats travel with every cell. The smallest cells are not findings:
\code{tools\_mcp}, at 4.7\% raw and 3.6\% deduplicated, is too thin a slice to read as a
rate. And the accuracy of \S\ref{sec:harness-arms} is established on 150
expert-labelled turns and nowhere else. The other 2{,}024 turns carry no reference, and
the labelled slate was stratified, not uniformly drawn, so that accuracy cannot be
carried onto this table.

Ticket analysis provides a complementary view of support demand. Across 2{,}211
ticket threads, 18\% of those flagged complete were not actually resolved, while
319 resolved threads were never flagged. The ticket system's status field is
therefore an unreliable proxy for outcome, supporting the need to re-triage
tickets from their content before routing follow-up work.

\subsection{RQ2: Does Human-Guided Training Improve the Diagnostic?}
\label{sec:harness-arms}
\label{sec:harness-ceiling}
\label{sec:harness-ledger}
\label{sec:harness-floor}

Sixteen harness arms were evaluated against the frozen expert reference: three
were accepted, nine were rejected, and four were reference arms (the baseline,
a null control and two replication draws). Table~\ref{tab:ledger} reports the
accepted path. Agreement rises from 108/150 to 123/150 turns. On the
session-disjoint evaluation split, it rises from 35/52 (67\%) to 40/52 (77\%).

\begin{table}[t]
\caption{The accepted harness changes, scored against the frozen expert
reference. Evaluation turns were not used to propose or select an arm.}
\label{tab:ledger}
\centering
\footnotesize
\begin{tabular}{@{}llccc@{}}
\toprule
Arm & Change & training (98) & eval. (52) & all \\
\midrule
--- & baseline, gate closed               & \textbf{73} & \textbf{35} & \textbf{108} \\
\textbf{A1} & rule-conditioned judge gate & 80 & 37 & 117 \\
\textbf{A2} & sufficiency $+$ clarification & 81 & 39 & 120 \\
\textbf{A3} & stricter clarification rules & \textbf{83} & \textbf{40} & \textbf{123} \\
\bottomrule
\end{tabular}
\end{table}

The first arm changed judge authority according to the rule that fired. It
corrected nine expert-labelled routes without introducing a regression. The
next two arms refined the sufficiency and clarification signals consumed by the
tree. The rejected arms generally attempted to change a judge's conclusion
without changing the evidence available to it. The accepted arms instead
changed how existing evidence was interpreted or combined.

Small aggregate changes require a null control. A semantically neutral edit
still changed 9 and 11 of the 171 scored routes in two replicates because
changing prompt bytes bypassed cached responses. Only two turns changed in both replicates, so
the unstable cases do not form a fixed subset that can be removed. We therefore
treat the null arm as a noise floor and accept a change only when its
pre-registered target cases improve while its controls remain stable.

\subsection{RQ3: What Does the Agent Judge Contribute?}
\label{sec:overturn}

\begin{figure}[t]
\centering
\resizebox{\columnwidth}{!}{%
\begin{tikzpicture}[x=1cm,y=1cm]
\node[anchor=east, font=\scriptsize, align=right] at (-0.12,1.04) {rule tree};
\node[anchor=east, font=\tiny, text=black!55] at (-0.12,0.83) {deterministic};
\filldraw[fill=rtNoFailF, draw=rtNoFailS, line width=0.5pt] (0.000,0.69) rectangle (1.932,1.19);
\node[font=\tiny, text=black!75] at (0.966,0.94) {26.3};
\filldraw[fill=rtRetrF, draw=rtRetrS, line width=0.5pt] (1.932,0.69) rectangle (3.389,1.19);
\node[font=\tiny, text=black!75] at (2.660,0.94) {19.8};
\filldraw[fill=rtKnowF, draw=rtKnowS, line width=0.5pt] (3.389,0.69) rectangle (4.473,1.19);
\node[font=\tiny, text=black!75] at (3.931,0.94) {14.8};
\filldraw[fill=rtPromptF, draw=rtPromptS, line width=0.5pt] (4.473,0.69) rectangle (5.425,1.19);
\node[font=\tiny, text=black!75] at (4.949,0.94) {13.0};
\filldraw[fill=rtModelF, draw=rtModelS, line width=0.5pt] (5.425,0.69) rectangle (5.964,1.19);
\node[font=\tiny, text=black!75] at (5.695,0.94) {7.3};
\filldraw[fill=rtToolsF, draw=rtToolsS, line width=0.5pt] (5.964,0.69) rectangle (6.313,1.19);
\filldraw[fill=rtUnkF, draw=rtUnkS, line width=0.5pt] (6.313,0.69) rectangle (7.350,1.19);
\node[font=\tiny, text=black!75] at (6.832,0.94) {14.1};
\node[anchor=east, font=\scriptsize, align=right] at (-0.12,0.19) {\textbf{agent judge}};
\node[anchor=east, font=\tiny, text=black!55] at (-0.12,-0.02) {probes, then commits};
\filldraw[fill=rtNoFailF, draw=rtNoFailS, line width=0.5pt] (0.000,-0.16) rectangle (2.179,0.34);
\node[font=\tiny, text=black!75] at (1.089,0.09) {29.6};
\filldraw[fill=rtRetrF, draw=rtRetrS, line width=0.5pt] (2.179,-0.16) rectangle (3.772,0.34);
\node[font=\tiny, text=black!75] at (2.975,0.09) {21.7};
\filldraw[fill=rtKnowF, draw=rtKnowS, line width=0.5pt] (3.772,-0.16) rectangle (5.029,0.34);
\node[font=\tiny, text=black!75] at (4.400,0.09) {17.1};
\filldraw[fill=rtPromptF, draw=rtPromptS, line width=0.5pt] (5.029,-0.16) rectangle (5.964,0.34);
\node[font=\tiny, text=black!75] at (5.496,0.09) {12.7};
\filldraw[fill=rtModelF, draw=rtModelS, line width=0.5pt] (5.964,-0.16) rectangle (6.252,0.34);
\filldraw[fill=rtToolsF, draw=rtToolsS, line width=0.5pt] (6.252,-0.16) rectangle (6.655,0.34);
\filldraw[fill=rtUnkF, draw=rtUnkS, line width=0.5pt] (6.655,-0.16) rectangle (7.350,0.34);
\node[font=\tiny, text=black!75] at (7.003,0.09) {\textbf{9.5}};
\draw[-{Latex[length=1.4mm]}, thick, red!60!black] (6.98,0.63) -- (6.98,0.42);
\node[anchor=east, font=\tiny\itshape, text=red!60!black] at (6.88,0.52)
  {nothing decidable: 306 $\rightarrow$ 205};
\end{tikzpicture}%
}
\caption{The rule tree and the agent judge on the same 2{,}169 turns.
\textbf{Neither bar is the shipped label}: the judge overturns only under seven named
rules (\S\ref{sec:overturn}), so what ships lies between them. The disagreement is
concentrated on the two classes a support desk cares most about: turns blamed on the model,
and turns nobody could attribute.}
\label{fig:judgedist}
\end{figure}
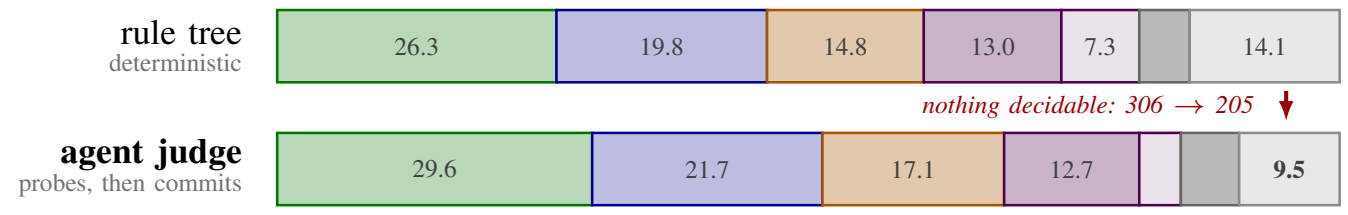

The rule tree and agent judge disagree on 354 of 2{,}169 turns (16.3\%), as
shown in Fig.~\ref{fig:judgedist}; they agree on the other 1{,}815 turns
(83.7\%). Where they differ, the judge assigns fewer turns to \code{model}
(159 versus 85) and leaves fewer unattributed (306 versus 205). The largest
flows are \code{model} $\rightarrow$ \code{no\_failure} (59), \code{unknown}
$\rightarrow$ \code{knowledge} (45), \code{prompt} $\rightarrow$
\code{no\_failure} (36), and \code{model} $\rightarrow$ \code{prompt} (31).
A confidence threshold did not provide an effective override policy. The
reported confidence represents the type and amount of evidence available, not
a calibrated probability that the proposed route is correct; it was often
highest when the judge and tree already agreed.
Policy~C in Table~\ref{tab:gates} exhibits this limitation, and a later rubric
revision was reverted when evaluation agreement fell from 72\% to 66\%.

The rule that produced the tree's route was more informative. On the 98
training turns, the judge alone was correct in ten cases and the tree alone in
seven. The judge was most useful when the tree inferred a route from missing or
weak evidence, because the judge could probe the corpus for additional
evidence. It was less reliable when the tree's route followed positive evidence
that the judge could not independently improve.

We therefore use a deterministic, rule-conditioned override gate. In
Table~\ref{tab:gates}, every override policy except the confidence-only gate
improves agreement by 7--9 percentage points on both splits relative to the
tree alone.
The evaluation split does not distinguish the three best policies, and the
ungated judge produces fewer spurious \code{knowledge} routes than the shipped
policy. Rule conditioning was selected for operational reasons: its seven
conditions are inspectable, its behavior is predictable if judge quality
degrades, and the tree bounds its failure modes. On the complete expert
reference, this gate corrects nine routes without a regression. These results
support using the judge, but do not establish that rule conditioning is more
accurate than accepting every judge proposal.

\begin{table}[t]
\caption{Gate policies on the same frozen records, measured earlier in the campaign, hence
$n{=}99/45$. ``Spurious \code{knowledge}'' counts turns wrongly sent to documentation
authoring. Bold marks the best value in a column. D is the policy we ship.}
\label{tab:gates}
\centering
\resizebox{\columnwidth}{!}{%
\footnotesize
\begin{tabular}{@{}llccc@{}}
\toprule
 & policy & train (99) & eval (45) & spurious \code{knowledge} \\
\midrule
A & rule tree alone, no judge           & 73\% & 64\% & 5 \\
B & judge alone, no gate                & 80\% & \textbf{71\%} & \textbf{1} \\
C & judge if confidence $\ge0.8$        & 77\% & 67\% & 2 \\
D & judge if the tree fired a weak rule \emph{(shipped)} & 81\% & \textbf{71\%} & 5 \\
E & D, and confidence $\ge0.7$          & \textbf{82\%} & \textbf{71\%} & 5 \\
\bottomrule
\end{tabular}%
}
\end{table}

The override gate reuses cached judge proposals and is therefore deterministic;
every pre-registered target must change as expected. Its authority is limited
to routing. Ticket replies, corpus changes, and pull requests remain subject to
expert review regardless of which component assigns the route.

% =====================================================================

\subsection{RQ4: Do Diagnoses Lead to Effective Repairs?}
\label{sec:production}
% =====================================================================

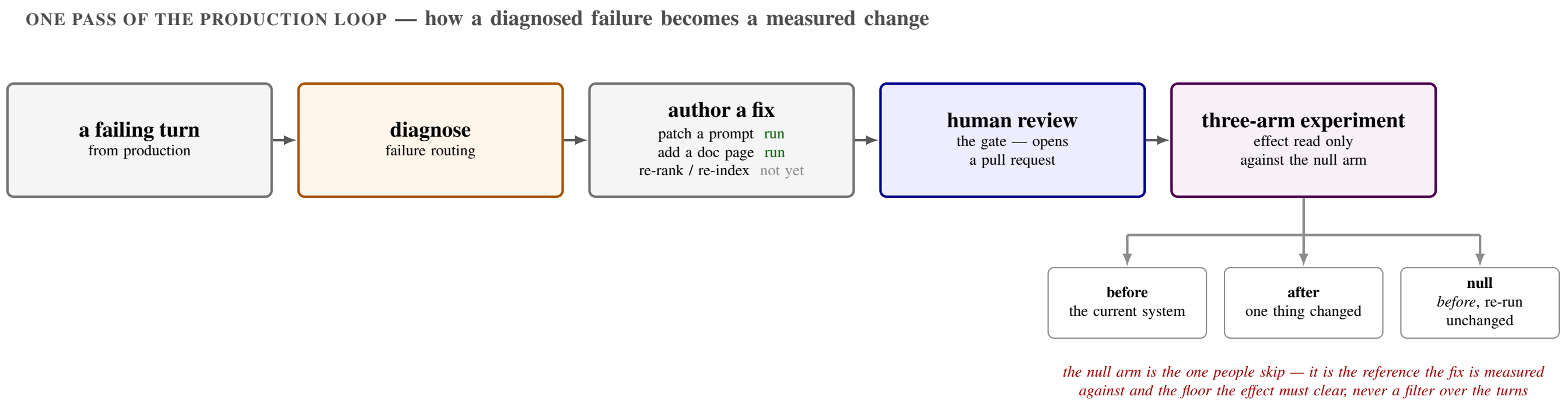
\begin{figure*}[t]
\centering
\begin{tikzpicture}[
  font=\tiny,
  st/.style={draw=black!55, thick, rounded corners=2pt, fill=black!4,
             align=center, inner sep=3.5pt, anchor=north,
             text width=2.75cm, minimum height=1.28cm},
  diag/.style={st, draw=orange!65!black, fill=orange!8},
  hgate/.style={st, draw=blue!55!black, fill=blue!7},
  meas/.style={st, draw=violet!60!black, fill=violet!6},
  arm/.style={draw=black!45, rounded corners=2pt, fill=white, minimum height=0.80cm,
              align=center, inner sep=2.5pt, anchor=north, text width=1.62cm},
  ar/.style={-{Latex[length=1.5mm]}, thick, black!65},
  hdr/.style={font=\scriptsize\bfseries, anchor=west}
]
\node[hdr, text=black!70] at (0,1.66) {\textsc{one pass of the production loop} --- how a diagnosed failure becomes a measured change};
\node[st]     (t) at (1.42,0.95) {{\scriptsize\bfseries a failing turn}\\from production};
\node[diag]   (d) at (4.72,0.95) {{\scriptsize\bfseries diagnose}\\failure routing};
\node[st]     (f) at (8.02,0.95) {{\scriptsize\bfseries author a fix}\\[1pt]
  patch a prompt \ \textcolor{green!40!black}{run}\\
  add a doc page \ \textcolor{green!40!black}{run}\\
  re-rank / re-index \ \textcolor{black!45}{not yet}};
\node[hgate]  (h) at (11.32,0.95) {{\scriptsize\bfseries human review}\\the gate --- opens\\a pull request};
\node[meas]   (m) at (14.62,0.95) {{\scriptsize\bfseries three-arm experiment}\\effect read only\\against the null arm};
\foreach \a/\b in {t/d, d/f, f/h, h/m} { \draw[ar] (\a.east) -- (\b.west); }

% One bus under the experiment box, then equal-length drops into the three arms.
\coordinate (abus) at ($(m.south)+(0,-0.42)$);
\node[arm] (a1) at (12.62,-1.14) {\textbf{before}\\the current system};
\node[arm] (a2) at (14.62,-1.14) {\textbf{after}\\one thing changed};
\node[arm] (a3) at (16.62,-1.14) {\textbf{null}\\\emph{before}, re-run\\unchanged};
\draw[black!45, thick] (m.south) -- (abus);
\draw[black!45, thick] (a1.north |- abus) -- (a3.north |- abus);
\foreach \a in {a1,a2,a3} { \draw[ar, black!45] (\a.north |- abus) -- (\a.north); }
\node[font=\tiny\itshape, text=red!60!black, anchor=north, text width=6.6cm, align=center]
  at (14.62,-2.14) {the null arm is the one people skip --- it is the reference the fix is
  measured against and the floor the effect must clear, never a filter over the turns};
\end{tikzpicture}
\caption{One pass of the production loop. Only the first two fix kinds have been run; a
retrieval-side repair is designed and not yet exercised. An effect is read as \emph{fix
against null}, never \emph{fix against before}: selecting a target set on a stochastic
symptom makes the before-arm rate 1.0 by construction. Numbers are in
\S\ref{sec:production}.}
\label{fig:prodloop}
\end{figure*}

The final evaluation tests whether a diagnosed failure can guide a targeted
repair. We examine one prompt failure and one knowledge gap. For each case, the
diagnostic identifies the responsible component, a candidate change is
authored, and its effect is measured by replaying production questions
(Fig.~\ref{fig:prodloop}). These are controlled offline experiments; neither
change was automatically promoted to the deployed system.

Each experiment has three arms: \emph{before} uses the current system,
\emph{after} changes exactly one component, and \emph{null} repeats the current
system with the generation cache bypassed. The null arm measures stochastic
variation: in these experiments, an unchanged configuration altered roughly
one quarter of individual responses. Effects are therefore measured between
the after and null arms, which are independent draws made after target cases
have been selected. Comparing after directly with the selection baseline would
overestimate improvement through regression to the mean.

We require an observed effect to exceed both the change in aggregate rate and
the number of per-turn flips in the null arm. This distinction matters because
opposing changes may cancel in aggregate; in one run, the rate changed by only
$+0.016$ although 27 of 188 turns (14.4\%) changed. The primary metrics below
are deterministic string matches, so the repair results do not depend on the
agent judge evaluated in RQ3. Retrieved contexts are frozen for the prompt
experiment. For the knowledge experiment, the candidate document is injected
directly, isolating its potential contribution while intentionally bypassing
retrieval.

\subsubsection{Questions and Reference Answers}
\label{sec:questionset}

Both experiments use a frozen question set derived from \FacAdj{} traffic.
This choice preserves requests that the existing corpus cannot answer, whereas
questions synthesized from a corpus tend to represent information already
present in that corpus~\cite{rageval2024,yourbench2025}. From 3{,}495 session
snapshots, 2{,}320 user messages were deduplicated and consolidated into
\textbf{1{,}512 canonical questions}. A bounded-tool agent drafted reference
answers using only sources it opened, after which a human reviewer accepted,
edited, or rejected them. Before experimentation, we content-hashed two subsets:
288 questions for the prompt repair and 68 for the knowledge repair.

\subsubsection{Repairing a Prompt Failure}
\label{sec:case-prompt}

Harness analysis identified over-clarification as the largest cluster assigned
to \code{prompt}: the assistant asks a follow-up question even when retrieved
evidence is sufficient. We evaluated a change to the conversation prompt used
by 99.5\% of traffic.

Across 288 frozen questions and two generation models, the change significantly
reduced clarifying questions relative to the null arm (Table~\ref{tab:clartradeoff}).
Responses became longer and included more citations, indicating that the
reduction was not caused by truncation. The change also increased
under-clarification---cases in which the assistant should have requested missing
system information. Beneficial changes outnumbered adverse changes by about
four to one, but the tradeoff requires expert judgment before deployment. The
control set also changed, so this experiment cannot distinguish the intended
instruction from the effect of modifying the prompt more generally; a placebo
prompt is required to isolate that mechanism.

\begin{table}[t]
\centering
\caption{Turn counts over all 288 questions. The three rows are mutually exclusive and sum
to 288. Row~1 counts every clarifying question asked, necessary or not.
Under-clarification is the opposite defect and is necessity-aware: a question naming no
system that the assistant answered instead of asking about. The effect is read
null against fix. $p$ is exact McNemar on the turns that changed.}
\label{tab:clartradeoff}
\footnotesize
\setlength{\tabcolsep}{4pt}
\begin{tabular}{@{}lccc@{\hskip 9pt}ccc@{}}
\toprule
 & \multicolumn{3}{c}{\code{claudeopus48}} & \multicolumn{3}{c}{\code{gpt5.6}} \\
\cmidrule(lr){2-4}\cmidrule(l){5-7}
turns, of 288 & before & null & \textbf{fix} & before & null & \textbf{fix} \\
\midrule
clarifying question asked & 145 & 128 & \textbf{101} & 69 & 68 & \textbf{37} \\
\textbf{under}-clarification & 3 & 4 & 9 & 15 & 17 & 24 \\
neither defect & 140 & 156 & 178 & 204 & 203 & 227 \\
\midrule
net, null $\rightarrow$ fix & \multicolumn{3}{c}{$-27$ asked\,/\,$+5$ under}
                            & \multicolumn{3}{c}{$-31$ asked\,/\,$+7$ under} \\
$p$ (asked\,/\,under) & \multicolumn{3}{c}{0.006\,/\,0.18}
                      & \multicolumn{3}{c}{$9.3{\times}10^{-6}$\,/\,0.039} \\
\bottomrule
\end{tabular}
\end{table}

\subsubsection{Repairing a Knowledge Gap}
\label{sec:case-knowledge}

The router also identified questions about AI coding assistants for which a
corpus probe found no coverage of five named tools. A documentation page was
drafted and approved by the content owner. Directly injecting this page raised
fact utilization from \textbf{0.070} to \textbf{0.202}, compared with
\textbf{0.060} when a same-sized unrelated page was injected. This result
isolates the value of the missing content, but represents an upper bound rather
than an end-to-end repair because it bypasses retrieval and re-indexing.

When the new page was chunked into the offline scoring corpus,
Table~\ref{tab:kfixpanel} showed higher completeness and faithfulness on target
questions and little change on controls for both generation models. Correctness
also rises above its null arm, but we do not claim it: unlike the other two
dimensions it re-retrieves from the scoring corpus, and its null moved 0.072 on
one model where theirs moved under 0.01.

The utilization increase supports the \code{knowledge} attribution for this
cluster. Across the broader question set, however, \textbf{76\% of gold facts
missing from retrieved context were already present in the corpus}; their
absence was therefore a retrieval failure. The remaining 24\% is an upper bound
on genuine documentation gaps, since the offline snapshot is a subset of the
production index. This distinction directs support effort toward either
retrieval improvement or content authoring rather than treating every missing
fact as a request for new documentation.

\begin{table}[t]
\centering
\caption{Target is the four questions the new page answers, control the other 64 it cannot
reach. The null arm is the before arm re-run through a cold cache. The judge is
cross-family to both generation arms and was chosen before any result.}
\label{tab:kfixpanel}
\footnotesize
\setlength{\tabcolsep}{5pt}
\begin{tabular}{llccc@{\hskip 12pt}ccc}
\toprule
 & & \multicolumn{3}{c}{\textsc{target} ($n{=}4$)} & \multicolumn{3}{c}{\textsc{control} ($n{=}64$)} \\
\cmidrule(lr){3-5}\cmidrule(l){6-8}
dimension & gen & before & \textbf{after} & null & before & after & null \\
\midrule
completeness & opus & 0.118 & \textbf{0.299} & 0.127 & 0.437 & 0.411 & 0.447 \\
completeness & gpt  & 0.098 & \textbf{0.173} & 0.089 & 0.348 & 0.370 & 0.365 \\
\addlinespace
faithfulness & opus & 0.691 & \textbf{0.857} & 0.695 & 0.716 & 0.734 & 0.721 \\
faithfulness & gpt  & 0.594 & \textbf{0.762} & 0.512 & 0.613 & 0.614 & 0.586 \\
\addlinespace
correctness  & opus & 0.700 & 0.773 & 0.722 & 0.622 & 0.638 & 0.617 \\
correctness  & gpt  & 0.500 & 0.583 & 0.428 & 0.487 & 0.471 & 0.491 \\
\bottomrule
\end{tabular}
\end{table}

\section{Discussion and Limitations}
\label{sec:discussion}
% =====================================================================

Three design lessons are likely to transfer to other facilities. First,
retrieval should be evaluated at both the production-result and corpus levels;
the distinction determines whether a retrieval engineer or documentation owner
can act on the failure. Second, answer-quality scores should not be treated as
repair decisions. A visible symptom does not uniquely identify the component
that caused it, and the reliability of attribution is bounded by the evidence
detectors that feed the policy. Third, a repair evaluation should measure both
the target behavior and its inverse. The prompt repair reduced unnecessary
clarification but also increased under-clarification, a tradeoff that a
single-direction metric would miss.

The study has several limitations. All labelled data come from \Fac{},
and approximately two thirds of the 150-turn reference received only one expert
label. The class distribution is unbalanced, making per-route estimates
unreliable for rare \code{model}, \code{unknown}, and \code{tools\_mcp} cases.
We report exact agreement without a chance-corrected statistic, and the held-out
split contains only 52 turns. The gate-policy comparison in
Table~\ref{tab:gates} is the only agentic-versus-deterministic control. Finally,
each repair study represents one experimental campaign. Null arms estimate
run-to-run variation but do not substitute for repeated independent trials.

The implemented system also stops short of fully autonomous repair. Three of
the five failure routes have not been exercised through an end-to-end repair
path, and no generated knowledge change has been merged and re-indexed in the
deployed corpus. The remaining errors, rejected harness arms, and finding that
76\% of missing retrieved facts already exist in the corpus all identify
retrieval quality as the highest-priority next step.

\section{Conclusion}
\label{sec:conclusion}
% =====================================================================

Deploying a RAG assistant creates a corresponding operational challenge:
determining why an answer failed and who can repair it. \sys{} addresses this
challenge by evaluating answer quality separately from retrieval sufficiency
and by testing retrieval at both the production-result and corpus levels. This
distinction separates missing documentation from failed retrieval and converts
quality evidence into a component-level work item.

Across 2{,}174 production turns, the router assigned a failure mode to 1{,}295
turns and explicitly left 309 unattributed. Human-guided harness training
increased agreement with a 150-turn expert reference from 108 to 123 turns,
including an increase from 67\% to 77\% on a session-disjoint evaluation split.
An agent judge provided useful independent evidence when the rule tree reasoned
from missing information; a rule-conditioned gate bounded its authority more
effectively than a confidence threshold.

The expert-review interface is central to this improvement process. It produces
the reference labels used for evaluation and captures rationales that motivate
bounded changes to rules, prompts, and knowledge. Controlled repair studies
showed that diagnoses can guide prompt and documentation changes, while also
revealing adverse effects that require expert judgment.

The system therefore demonstrates a governed path from support interactions to
measured improvement, but not a fully closed autonomous loop. Candidate repairs
have been evaluated offline; generated knowledge changes have not yet been
merged and re-indexed in production. Completing and evaluating that final stage
is future work.

\section*{Acknowledgment}
This research used resources of the Argonne Leadership Computing Facility, a U.S. Department of
Energy (DOE) Office of Science user facility at Argonne National Laboratory (ANL) operated under
Contract No. DE-AC02-06CH11357.
J.J., A.B., K.R., M.K., Y.G., H.S.S., V.V., M.E.P., and H.Z. were supported by the Office of Science,
U.S. Department of Energy, under contract DE-AC02-06CH11357.

% IEEEtran balances the final bibliography page naturally with this author layout.
\bibliographystyle{IEEEtran}
\bibliography{references}

\end{document}